\documentclass[10pt,twocolumn,letterpaper]{article}

\usepackage{iccv}
\usepackage{times}
\usepackage{epsfig}
\usepackage{graphicx}
\usepackage{amsmath}
\usepackage{amssymb}

\usepackage{booktabs}
\usepackage{xcolor}
\usepackage{amsmath, bm, subfigure, epstopdf, url, pifont, overpic, cases}
\usepackage{latexsym, amssymb, bbding, multirow, makecell,  diagbox, enumitem, caption}
\usepackage{array, enumitem, soul, algorithm, algpseudocode}
\newcommand{\R}[1]{\textcolor[rgb]{1.00,0.00,0.00}{#1}}
\newcommand{\B}[1]{\textcolor[rgb]{0.00,0.00,1.00}{#1}}

\usepackage{arydshln}
\newcommand{\algname}{SwinIR}
\usepackage{multirow, multicol}
\usepackage{adjustbox}

\newcommand{\name}{0}
\newcommand{\h}{0}
\newcommand{\w}{0.15}
\newcommand{\wa}{0.15}
\newlength \g

\usepackage[pagebackref=true,breaklinks=true,letterpaper=true,colorlinks,bookmarks=false]{hyperref}

\iccvfinalcopy %

\def\iccvPaperID{10} %

\begin{document}

\title{SwinIR: Image Restoration Using Swin Transformer}
\author{\hspace{-0.4cm}Jingyun Liang$^{1}$ ~ Jiezhang Cao$^{1}$  ~ Guolei Sun$^{1}$ ~ Kai Zhang$^{1,}$\thanks{Corresponding author.} ~ Luc Van Gool$^{1,2}$ ~ Radu Timofte$^{1}$\\
$^{1}$Computer Vision Lab, ETH Zurich, Switzerland\quad\quad $^{2}$KU Leuven, Belgium\\
{\tt\small \hspace{-0.6cm}\{jinliang, jiezcao, guosun, kai.zhang, vangool, timofter\}@vision.ee.ethz.ch}\\ 
{\tt\small }\url{https://github.com/JingyunLiang/SwinIR}
}

\maketitle

\begin{abstract}
Image restoration is a long-standing low-level vision problem that aims to restore high-quality images from low-quality images (\eg, downscaled, noisy and compressed images). While state-of-the-art image restoration methods are based on convolutional neural networks, few attempts have been made with Transformers which show impressive performance on high-level vision tasks. In this paper, we propose a strong baseline model SwinIR for image restoration based on the Swin Transformer. SwinIR consists of three parts: shallow feature extraction, deep feature extraction and high-quality image reconstruction. In particular, the deep feature extraction module is composed of several residual Swin Transformer blocks (RSTB), each of which has several Swin Transformer layers together with a residual connection. We conduct experiments on three representative tasks: image super-resolution (including classical, lightweight and real-world image super-resolution), image denoising (including grayscale and color image denoising) and JPEG compression artifact reduction. Experimental results demonstrate that SwinIR outperforms state-of-the-art methods on different tasks by $\textbf{up to 0.14$\sim$0.45dB}$, while the total number of parameters can be reduced by $\textbf{up to 67\%}$.
\end{abstract}

\section{Introduction}
Image restoration, such as image super-resolution (SR), image denoising and JPEG compression artifact reduction, aims to reconstruct the high-quality clean image from its low-quality degraded counterpart. Since several revolutionary work~\cite{dong2014srcnn, kim2016vdsr, zhang2017DnCNN, zhang2017IRCNN}, convolutional neural networks (CNN) have become the primary workhorse for image restoration~\cite{ledig2017srresnet, lim2017edsr, ledig2017srresnet, wang2018esrgan, zhang2018ffdnet, zhang2018rcan, fritsche2019frequency, zhang2018srmd, li2019SRFBN, kai2021bsrgan, zhang2021DPIR}. 

Most CNN-based methods focus on elaborate architecture designs such as residual learning~\cite{ledig2017srresnet, lim2017edsr} and dense connections~\cite{zhang2018RDN, wang2018esrgan}. Although the performance is significantly improved compared with traditional model-based methods~\cite{timofte2014a, dabov2007bm3d,gu2012fast}, they generally suffer from two basic problems that stem from the basic convolution layer. First, the interactions between images and convolution kernels are content-independent. Using the same convolution kernel to restore different image regions may not be the best choice. Second, under the principle of local processing, convolution is not effective for long-range dependency modelling.%

As an alternative to CNN, Transformer~\cite{vaswani2017transformer} designs a self-attention mechanism to capture global interactions between contexts and has shown promising performance in several vision problems~\cite{carion2020DETR, touvron2020DeiT, dosovitskiy2020ViT, liu2021swin}. 
However, vision Transformers for image restoration~\cite{chen2021IPT, cao2021videosr} usually divide the input image into patches with fixed size (\eg, 48$\times$48) and process each patch independently. Such a strategy inevitably gives rise to two drawbacks. First, border pixels cannot utilize neighbouring pixels that are out of the patch for image restoration. Second, the restored image may introduce border artifacts around each patch. While this problem can be alleviated by patch overlapping, it would introduce extra computational burden.

\begin{figure}[!tbp]
\captionsetup{font=small}%
\begin{center}
\vspace{-0mm}
\begin{overpic}[width=7.cm]{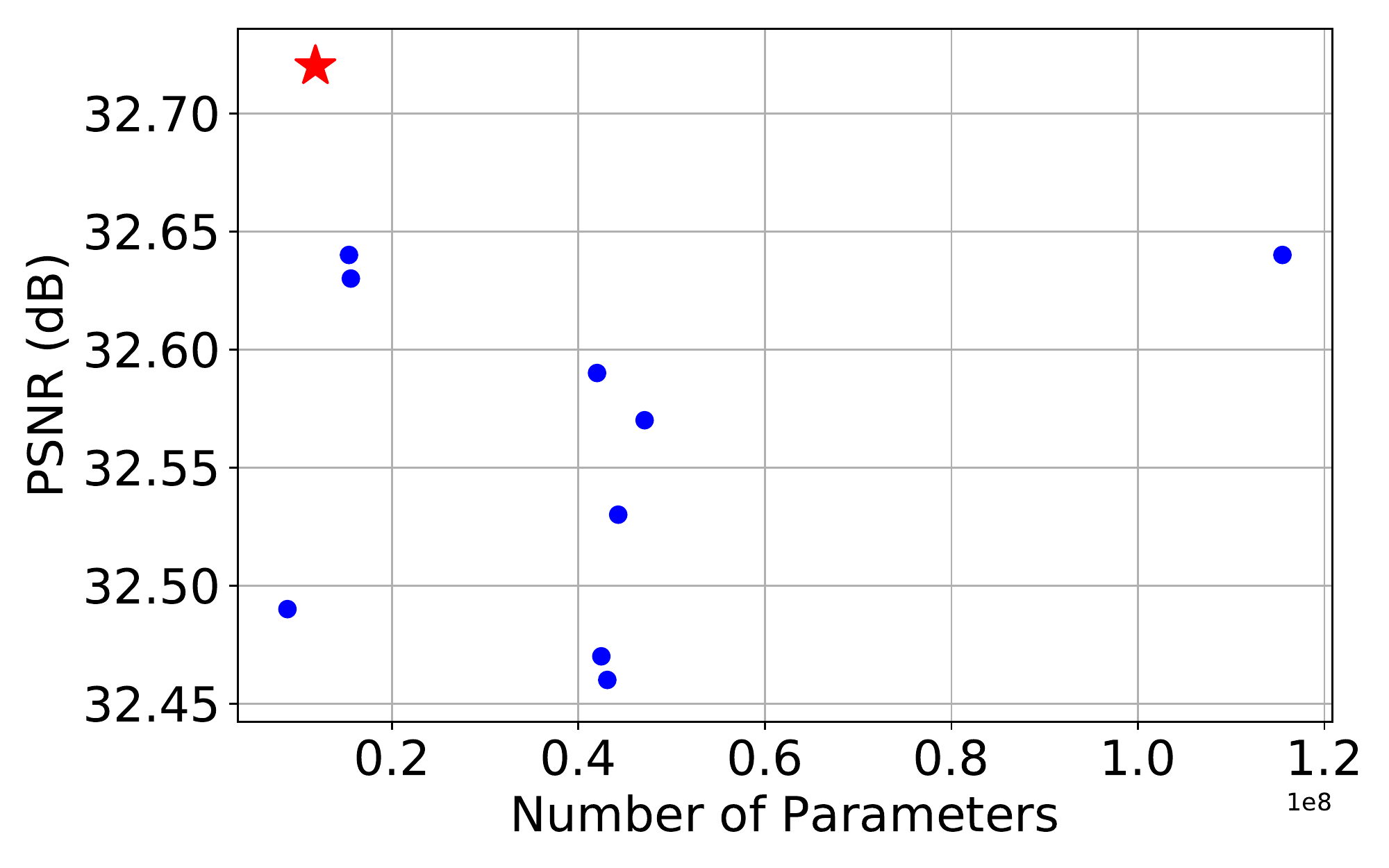}
\put(44.5,11.8){\color{black}{ \fontsize{6}{5}\selectfont{ {\makecell{EDSR  (CVPR2017)}}}}}
\put(21,17.5){\color{black}{ \fontsize{6}{5}\selectfont{ {\makecell{RNAN (ICLR2019)}}}}}
\put(44.5,24){\color{black}{ \fontsize{6}{5}\selectfont{ {\makecell{OISR (CVPR2019)}}}}}
\put(43.6,15){\color{black}{ \fontsize{6}{5}\selectfont{ {\makecell{RDN (CVPR2018)}}}}}
\put(25.7,40){\color{black}{ \fontsize{6}{5}\selectfont{ {\makecell{RCAN (ECCV2018)}}}}}
\put(47.3,30.8){\color{black}{ \fontsize{6}{5}\selectfont{ {\makecell{IGNN (NeurIPS2020)}}}}}
\put(25.6,43.3){\color{black}{ \fontsize{6}{5}\selectfont{ {\makecell{HAN (ECCV2020)}}}}}
\put(43.8,34.5){\color{black}{ \fontsize{6}{5}\selectfont{ {\makecell{NLSA (CVPR2021)}}}}}
\put(72,39.5){\color{black}{ \fontsize{6}{5}\selectfont{ {\makecell{IPT (CVPR2021)}}}}}
\put(24,57){\color{black}{ \fontsize{6}{5}\selectfont{ {\makecell{\textbf{SwinIR} (ours)}}}}}
\end{overpic}
\end{center}\vspace{-0.7cm}
\caption{PSNR results v.s the total number of parameters of different methods for image SR ($\times$4) on Set5~\cite{Set5}.}
\label{fig:para_psnr}
\vspace{-0.3cm}
\end{figure}
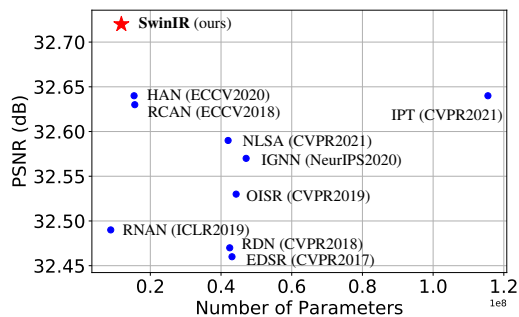

Recently, Swin Transformer~\cite{liu2021swin} has shown great promise as it integrates the advantages of both CNN and Transformer. On the one hand, it has the advantage of CNN to process image with large size due to the local attention mechanism. On the other hand, it has the advantage of Transformer to model long-range dependency with the shifted window scheme.

In this paper, we propose an image restoration model, namely SwinIR, based on Swin Transformer. More specifically, SwinIR consists of three modules: shallow feature extraction, deep feature extraction and high-quality image reconstruction modules. Shallow feature extraction module uses a convolution layer to extract shallow feature, which is directly transmitted to the reconstruction module so as to preserve low-frequency information. Deep feature extraction module is mainly composed of residual Swin Transformer blocks (RSTB), each of which utilizes several Swin Transformer layers for local attention and cross-window interaction. In addition, we add a convolution layer at the end of the block for feature enhancement and use a residual connection to provide a shortcut for feature aggregation. Finally, both shallow and deep features are fused in the reconstruction module for high-quality image reconstruction.

Compared with prevalent CNN-based image restoration models, Transformer-based SwinIR has several benefits: (1) content-based interactions between image content and attention weights, which can be interpreted as spatially varying convolution~\cite{cordonnier2019SAvsConv, elsayed2020revisiting, vaswani2021SAhaloing}. (2) long-range dependency modelling are enabled by the shifted window mechanism. (3) better performance with less parameters. For example, as shown in Fig.~\ref{fig:para_psnr}, SwinIR achieves better PSNR with less parameters compared with existing image SR methods.

\section{Related Work}
\subsection{Image Restoration}
Compared to traditional image restoration methods~\cite{gu2012fast, timofte2013anchored, timofte2014a, michaeli2013nonparametric, he2010darkchannel} which are generally model-based, learning-based methods, especially CNN-based methods, have become more popular due to their impressive performance. They often learn mappings between low-quality and high-quality images from large-scale paired datasets. Since pioneering work SRCNN~\cite{dong2014srcnn} (for image SR), DnCNN~\cite{zhang2017DnCNN} (for image denoising) and ARCNN~\cite{dong2015compression} (for JPEG compression artifact reduction), a flurry of CNN-based models have been proposed to improve model representation ability by using more elaborate neural network architecture designs, such as residual block~\cite{kim2016vdsr, cavigelli2017cas, zhang2021DPIR}, dense block~\cite{wang2018esrgan, zhang2018RDN, zhang2020RDNIR} and others~\cite{chen2016TNRD, lai2017LapSRN, zhang2018srmd, wang2019learning, wang2021unsupervised, wang2021learning, liang2021fkp, liang21hcflow, liang21manet, zhang2018ffdnet, tai2017memnet, isobe2020video, wei2021unsupervised, guo2020closed, cheng2021mfagan, deng2021deep, zhang2019RNAN, peng2019dsnet, jia2019focnet, fu2019jpeg, kim2019pseudo, fu2021model}. Some of them have exploited the attention mechanism inside the CNN framework, such as channel attention~\cite{zhang2018rcan, dai2019SAN, niu2020HAN}, non-local attention~\cite{liu2018NLRN, mei2021NLSA} and adaptive patch aggregation~\cite{zhou2020IGNN}.

\begin{figure*}[ht]
\captionsetup{font=small}%
\scriptsize
\begin{center}
\includegraphics[width=0.99\textwidth]{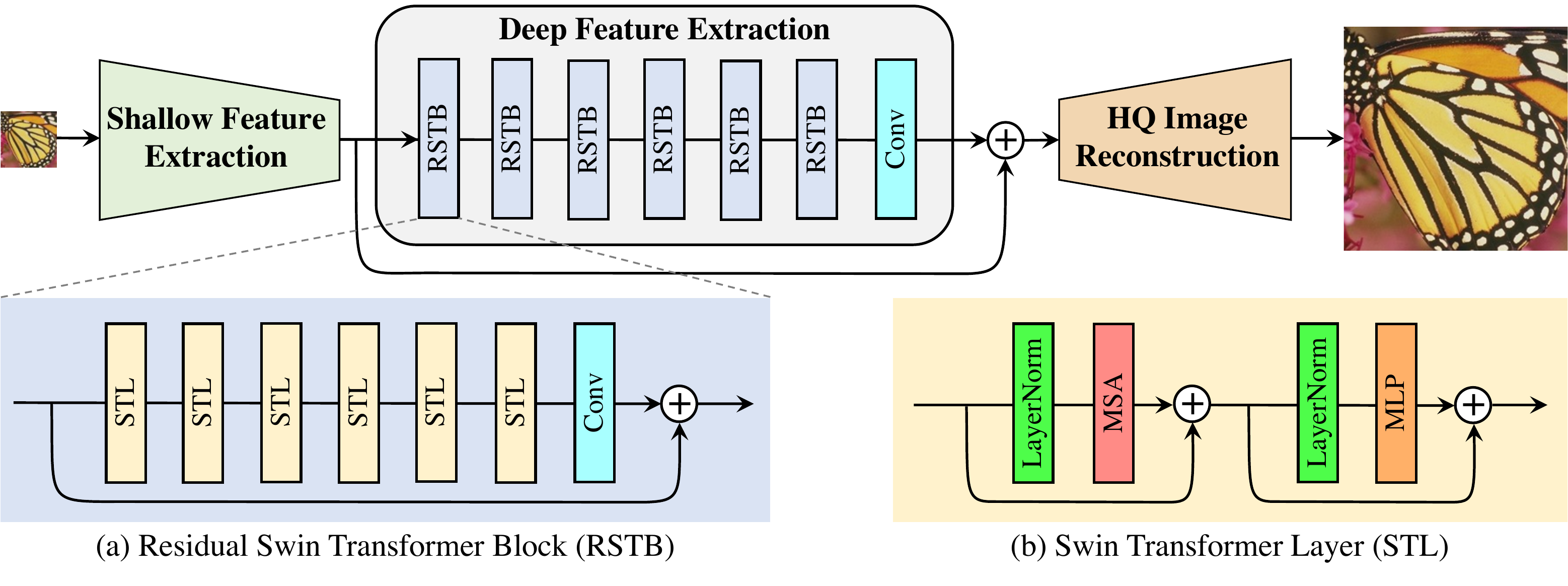}\label{fig:SwinIR_SR}%
\vspace{-0.3cm}
\caption{The architecture of the proposed SwinIR for image restoration.}\label{fig:architecture}
\end{center}
\vspace{-0.3cm}
\end{figure*}

\subsection{Vision Transformer}
Recently, natural language processing model Transformer~\cite{vaswani2017transformer} has gained much popularity in the computer vision community. When used in vision problems such as image classification~\cite{ramachandran2019SASA, dosovitskiy2020ViT, wu2020visual, liu2021swin, li2021localvit, liu2021transformer, vaswani2021SAhaloing}, object detection~\cite{carion2020DETR, liu2020deep, touvron2020DeiT, liu2021swin}, segmentation~\cite{wu2020visual, zheng2021rethinking, liu2021swin, cao2021swinunet} and crowd counting~\cite{liang2021transcrowd,sun2021boosting}, it learns to attend to important image regions by exploring the global interactions between different regions. 
Due to its impressive performance, Transformer has also been introduced for image restoration~\cite{chen2021IPT, cao2021videosr, wang2021uformer}.
Chen~\etal~\cite{chen2021IPT} proposed a backbone model IPT for various restoration problems based on the standard Transformer. However, IPT relies on large number of parameters (over 115.5M parameters), large-scale datasets (over 1.1M images) and multi-task learning for good performance. Cao~\etal~\cite{cao2021videosr} proposed VSR-Transformer that uses the self-attention mechanism for better feature fusion in video SR, but image features are still extracted from CNN. Besides, both IPT and VSR-Transformer are patch-wise attention, which may be improper for image restoration. In addition, a concurrent work~\cite{wang2021uformer} proposed a U-shaped architecture based on the Swin Transformer~\cite{liu2021swin}.

\section{Method}

\subsection{Network Architecture}
\label{sec:archi}
As shown in Fig.~\ref{fig:architecture}, SwinIR consists of three modules: shallow feature extraction, deep feature extraction and high-quality (HQ) image reconstruction modules. We employ the same feature extraction modules for all restoration tasks, but use different reconstruction modules for different tasks.

\vspace{-0.4cm}
\paragraph{Shallow and deep feature extraction.}
Given a low-quality (LQ) input $I_{\textit{LQ}}\in\mathbb{R}^{H\times W\times C_{in}}$ ($H$, $W$ and $C_{in}$ are the image height, width and input channel number, respectively), we use a $3\times 3$ convolutional layer $H_{\textit{SF}}(\cdot)$  to extract shallow feature $F_0\in\mathbb{R}^{H\times W\times C}$ as
\begin{equation}
F_0=H_{\textit{SF}}(I_{\textit{LQ}}),
\end{equation} 
where $C$ is the feature channel number. The convolution layer is good at early visual processing, leading to more stable optimization and better results~\cite{xiao2021early}. It also provides a simple way to map the input image space to a higher dimensional feature space. Then, we extract deep feature $F_{\textit{DF}}\in\mathbb{R}^{H\times W\times C}$ from $F_0$ as
\begin{equation}
F_{\textit{DF}}=H_{\textit{DF}}(F_0),
\end{equation} 
where $H_{\textit{DF}}(\cdot)$ is the deep feature extraction module and it contains $K$ {r}esidual {S}win {T}ransformer {b}locks (RSTB) and a $3\times 3$ convolutional layer. More specifically, intermediate features $F_{1},F_2, \ldots, F_K$ and the output deep feature $F_{\textit{DF}}$ are extracted block by block as
\begin{equation}
\begin{split}
F_{i}= H_{\textit{RSTB}_{i}}(F_{i-1}), \quad i=1, 2, \ldots, K,\\
F_{\textit{DF}}= H_{\textit{CONV}}(F_K),\qquad\qquad
\end{split}
\end{equation} 
where $H_{\textit{RSTB}_{i}}(\cdot)$ denotes the $i$-th RSTB and $H_{\textit{CONV}}$ is the last convolutional layer. Using a convolutional layer at the end of feature extraction can bring the inductive bias of the convolution operation into the Transformer-based network, and lay a better foundation for the later aggregation of shallow and deep features. 

\vspace{-0.4cm}
\paragraph{Image reconstruction.}
Taking image SR as an example, we reconstruct the high-quality image $I_{\textit{RHQ}}$ by aggregating shallow and deep features as
\begin{equation}
I_{\textit{RHQ}}=H_{\textit{REC}}(F_0+F_{\textit{DF}}),
\end{equation}
where $H_{\textit{REC}}(\cdot)$ is the function of the reconstruction module. Shallow feature mainly contain low-frequencies, while deep feature focus on recovering lost high-frequencies. With a long skip connection, SwinIR can transmit the low-frequency information directly to the reconstruction module, which can help deep feature extraction module focus on high-frequency information and stabilize training. For the implementation of reconstruction module, we use the sub-pixel convolution layer~\cite{shi2016subpixel} to upsample the feature. 

For tasks that do not need upsampling, such as image denoising and JPEG compression artifact reduction, a single convolution layer is used for reconstruction. Besides, we use residual learning to reconstruct the residual between the LQ and the HQ image instead of the HQ image. This is formulated as
\begin{equation}
I_{\textit{RHQ}} = H_{\textit{SwinIR}}(I_{\textit{LQ}}) +I_{\textit{LQ}},
\end{equation}
where $H_{\textit{SwinIR}}(\cdot)$ denotes the function of SwinIR.

\vspace{-0.4cm}
\paragraph{Loss function.}
For image SR, we optimize the parameters of SwinIR by minimizing the $L_1$ pixel loss
\begin{equation}
\mathcal{L}=\| I_{\textit{RHQ}}-I_{\textit{HQ}}\|_1,
\end{equation}
where $I_{\textit{RHQ}}$ is obtained by taking $I_{\textit{LQ}}$ as the input of SwinIR, and $I_{\textit{HQ}}$ is the corresponding ground-truth HQ image. For classical and lightweight image SR, we only use the naive $L_1$ pixel loss as same as previous work to show the effectiveness of the proposed network. For real-world image SR, we use a combination of pixel loss, GAN loss and perceptual loss~\cite{wang2018esrgan, kai2021bsrgan, wang2021realESRGAN, goodfellow2014GAN, johnson2016perceptual, wang2018esrgan} to improve visual quality.

For image denoising and JPEG compression artifact reduction, we use the Charbonnier loss~\cite{charbonnier1994Charbonnier}
\begin{equation}
\mathcal{L}= \sqrt{\|I_{\textit{RHQ}}-I_{\textit{HQ}}\|^2+\epsilon^2},%
\end{equation}
where $\epsilon$ is a constant that is empirically set to $10^{-3}$.

\subsection{Residual Swin Transformer Block}
As shown in Fig.~\ref{fig:architecture}\R{(a)}, the residual Swin Transformer block (RSTB) is a residual block with Swin Transformer layers (STL) and convolutional layers. Given the input feature $F_{i,0}$ of the $i$-th RSTB, we first extract intermediate features $F_{i,1},F_{i,2}, \ldots, F_{i,L}$ by $L$ Swin Transformer layers as
\begin{equation}
F_{i,j}=H_{\textit{STL}_{i,j}}(F_{i,j-1}), \quad j=1,2, \ldots, L,
\end{equation} 
where $H_{\textit{STL}_{i,j}}(\cdot)$ is the $j$-th Swin Transformer layer in the $i$-th RSTB. Then, we add a convolutional layer before the residual connection. The output of RSTB is formulated as
\begin{equation}
F_{i,out}=H_{\textit{CONV}_i}(F_{i,L})+F_{i,0},
\end{equation}
where $H_{\textit{CONV}_i}(\cdot)$ is the convolutional layer in the $i$-th RSTB. This design has two benefits. First, although Transformer can be viewed as a specific instantiation of spatially varying convolution~\cite{elsayed2020revisiting, vaswani2021SAhaloing}, covolutional layers with spatially invariant filters can enhance the translational equivariance of SwinIR. Second, the residual connection provides a identity-based connection from different blocks to the reconstruction module, allowing the aggregation of different levels of features.

\vspace{-0.4cm}
\paragraph{Swin Transformer layer.} Swin Transformer layer (STL)~\cite{liu2021swin} is based on the standard multi-head self-attention of the original Transformer layer~\cite{vaswani2017transformer}. The main differences lie in local attention and the shifted window mechanism. 
As shown in Fig.~\ref{fig:architecture}\R{(b)}, given an input of size ${H\times W\times C}$, Swin Transformer first reshapes the input to a $\frac{HW}{M^2}\times M^2\times C$ feature by partitioning the input into non-overlapping $M\times M$ local windows, where $\frac{HW}{M^2}$ is the total number of windows. Then, it computes the standard self-attention separately for each window (\ie, local attention). For a local window feature $X\in\mathbb{R}^{M^2\times C}$, the \textit{query}, \textit{key} and \textit{value} matrices $Q$, $K$ and $V$ are computed as
\begin{equation}
Q=XP_Q,\quad K=XP_K,\quad V=XP_V,
\end{equation}
where $P_Q$, $P_K$ and $P_V$ are projection matrices that are shared across different windows. Generally, we have $Q,K,V\in\mathbb{R}^{M^2\times d}$. The attention matrix is thus computed by the self-attention mechanism in a local window as
\begin{equation}
\text{Attention}(Q,K,V)=\text{SoftMax}(QK^T/\sqrt{d}+B)V,
\end{equation}
where $B$ is the learnable relative positional encoding. In practice, following \cite{vaswani2017transformer}, we perform the attention function for $h$ times in parallel and concatenate the results for multi-head self-attention (MSA). 

Next, a multi-layer perceptron (MLP) that has two fully-connected layers with GELU non-linearity between them is used for further feature transformations. The LayerNorm (LN) layer is added before both MSA and MLP, and the residual connection is employed for both modules. The whole process is formulated as
\begin{equation}
\begin{split}
X&=\text{MSA}(\text{LN}(X))+X,\\
X&=\text{MLP}(\text{LN}(X))+X.
\end{split}
\end{equation}

However, when the partition is fixed for different layers, there is no connection across local windows. Therefore, regular and shifted window partitioning are used alternately to enable cross-window connections \cite{liu2021swin}, where shifted window partitioning means shifting the feature by $(\lfloor\frac{M}{2}\rfloor, \lfloor\frac{M}{2}\rfloor)$ pixels before partitioning.

\subfigcapskip=-0.2cm
\begin{figure*}[!t]
\captionsetup{font=small}
\begin{center}
\subfigure[\hspace{-0.cm}]{\includegraphics[width=0.26\textwidth]{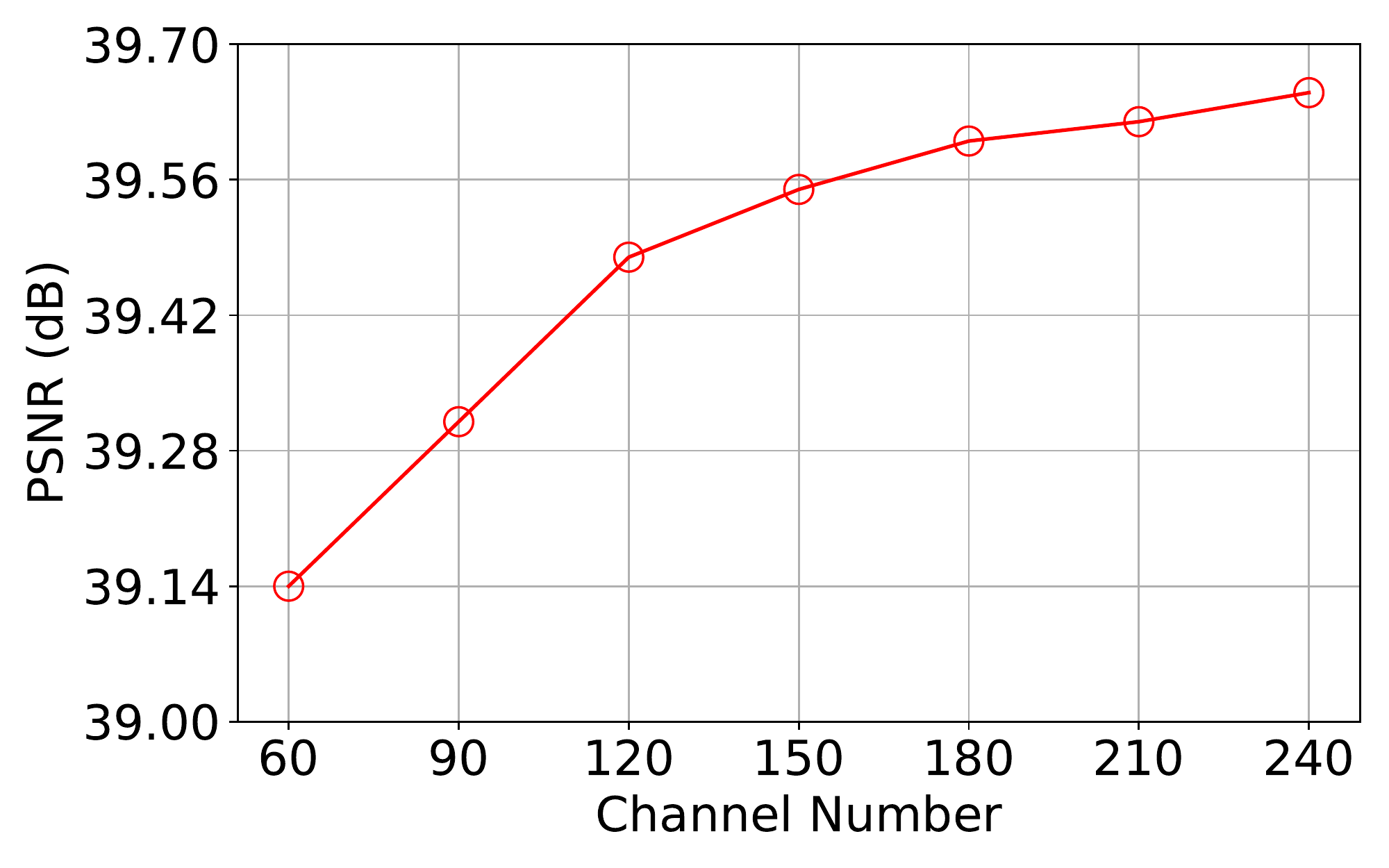}\label{fig:ablation_channelnumber}}\hspace{0.05\textwidth}\vspace{-0.1cm}
\subfigure[\hspace{-0.5cm}]{\includegraphics[width=0.26\textwidth]{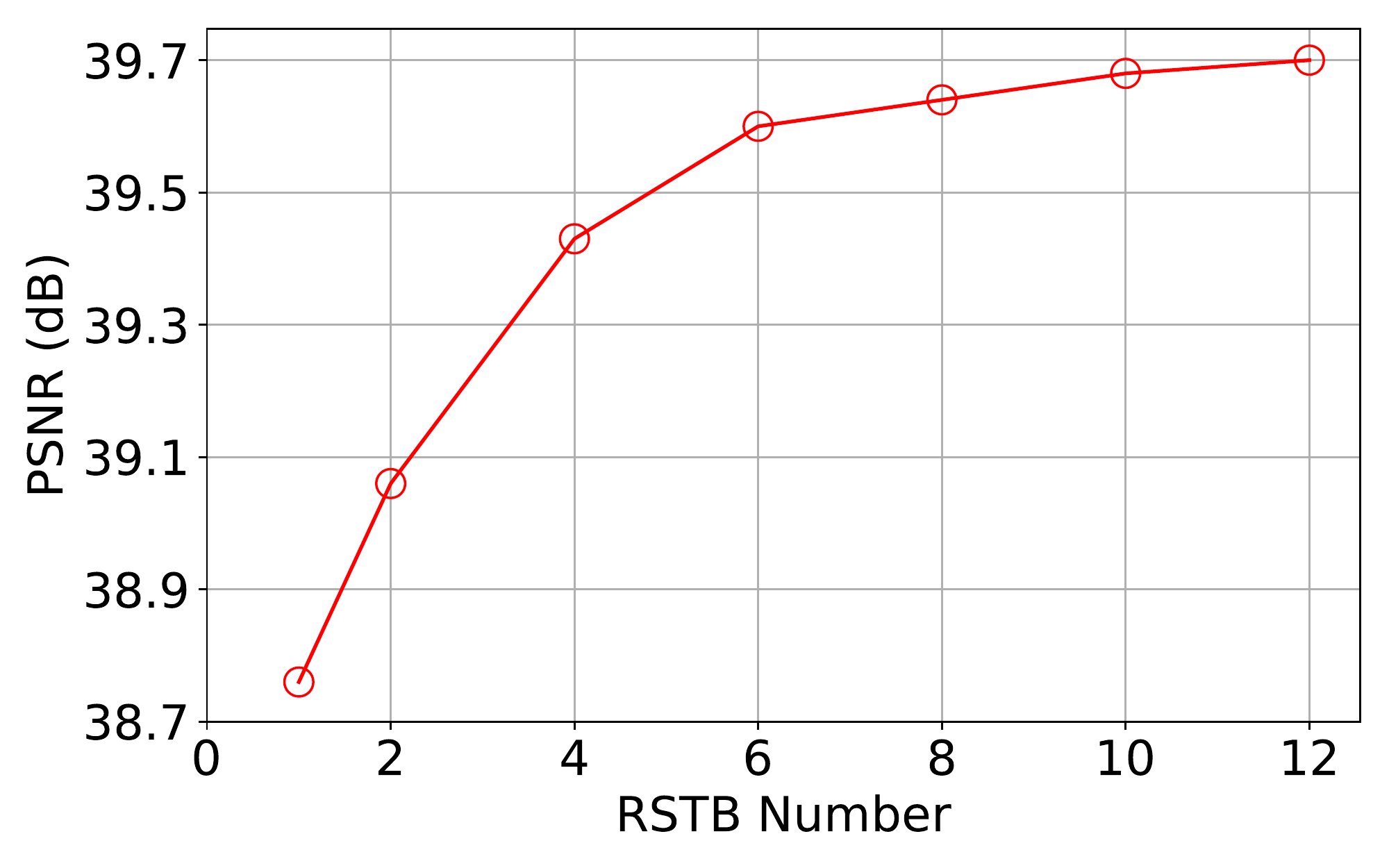}\label{fig:ablation_RSTBnumber}} \hspace{0.05\textwidth}\vspace{-0.1cm}
\subfigure[\hspace{-0.5cm}]{\includegraphics[width=0.26\textwidth]{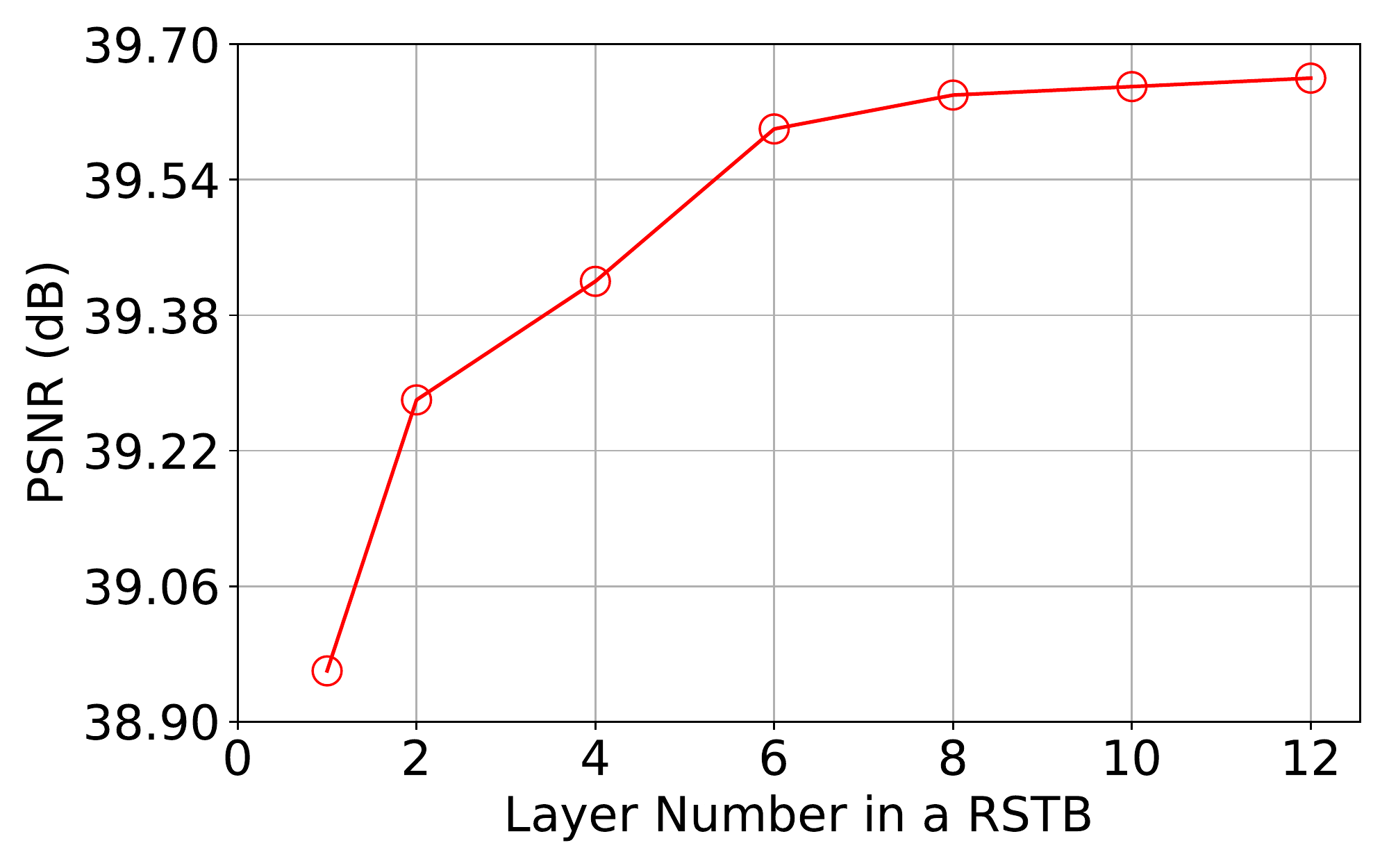}\label{fig:ablation_layernumber}} \vspace{-0.1cm}
\subfigure[\hspace{-0.cm}]{\includegraphics[width=0.26\textwidth]{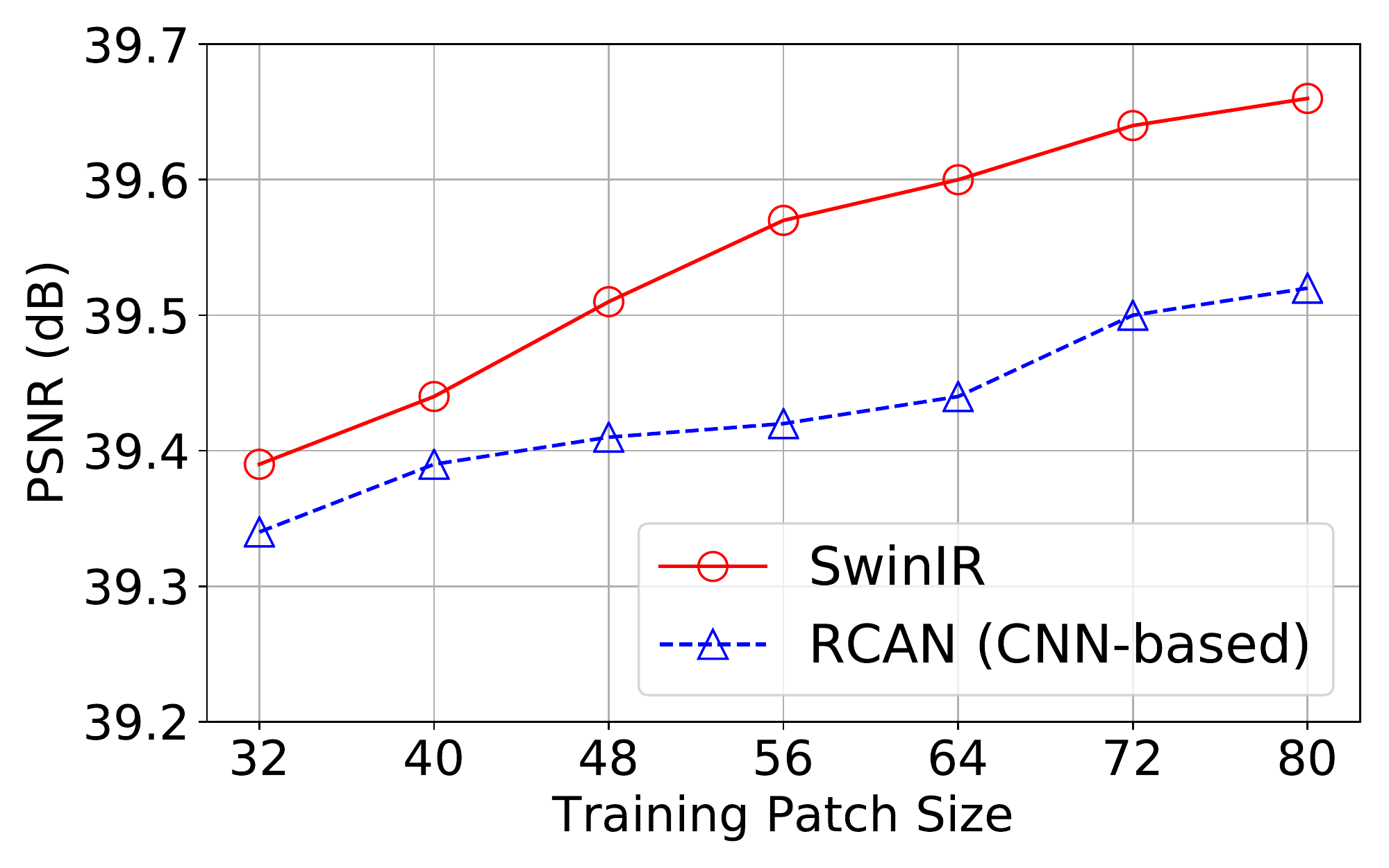}\label{fig:ablation_patchsize}}\hspace{0.05\textwidth}
\subfigure[\hspace{-0.5cm}]{\includegraphics[width=0.26\textwidth]{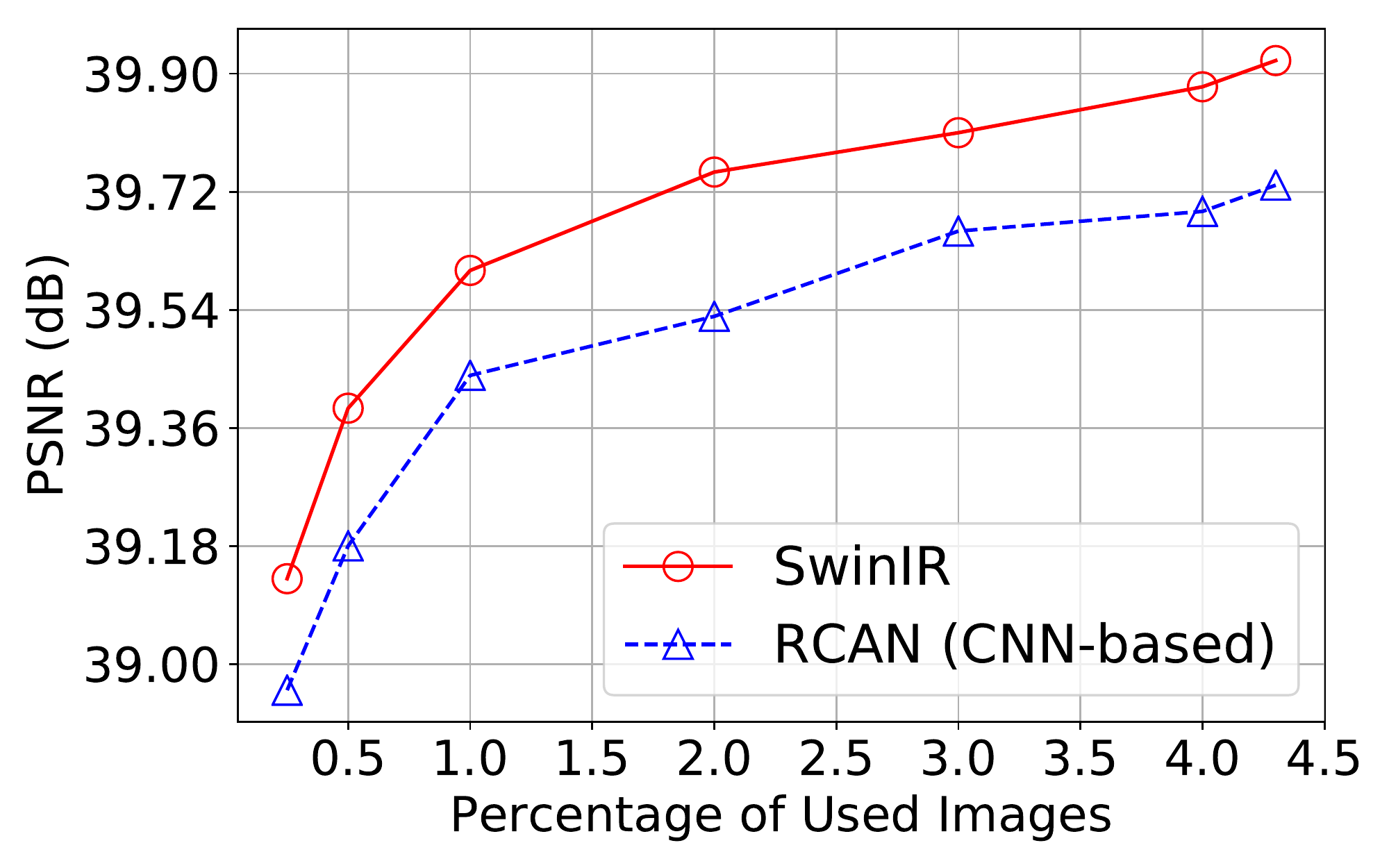}\label{fig:ablation_dataset}} \hspace{0.05\textwidth}
\subfigure[\hspace{-0.5cm}]{\includegraphics[width=0.26\textwidth]{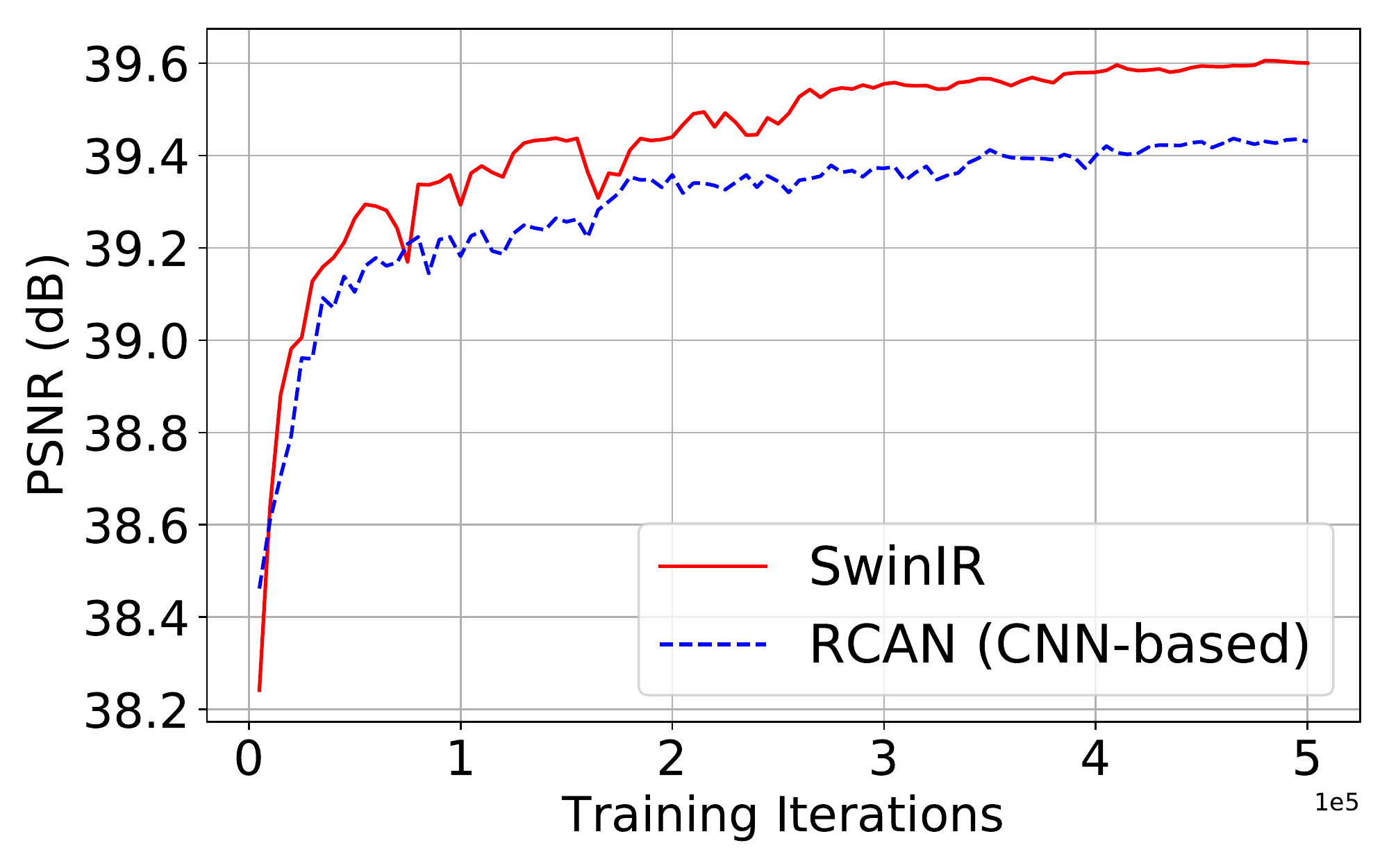}\label{fig:ablation_iteration}} 
\end{center}\vspace{-0.7cm}
\caption{Ablation study on different settings of SwinIR. Results are tested on Manga109~\cite{Manga109} for image SR ($\times 2$).}
\vspace{-0.3cm}
\label{fig:ablation}
\end{figure*}
\subfigcapskip=0.1cm

\section{Experiments}
\subsection{Experimental Setup}
For classical image SR, real-world image SR, image denoising and JPEG compression artifact reduction, the RSTB number, STL number, window size, channel number and attention head number are generally set to 6, 6, 8, 180 and 6, respectively. One exception is that the window size is set to 7 for JPEG compression artifact reduction, as we observe significant performance drop when using 8, possibly because JPEG encoding uses $8\times 8$ image partions. For lightweight image SR, we decrease RSTB number and channel number to 4 and 60, respectively. Following ~\cite{zhang2018rcan, niu2020HAN}, when self-ensemble strategy~\cite{lim2017edsr} is used in testing, we mark the model with a symbol ``+'', \eg, SwinIR+.  %
Due to page limit, training and evaluation details are provided in the supplementary.

\subsection{Ablation Study and Discussion}
For ablation study, we train SwinIR on DIV2K~\cite{DIV2K} for classical image SR ($\times 2$) and test it on Manga109~\cite{Manga109}.

\vspace{-0.4cm}
\paragraph{Impact of channel number, RSTB number and STL number.}
We show the effects of channel number, RSTB number and STL number in a RSTB on model performance in Figs.~\ref{fig:ablation_channelnumber},~\ref{fig:ablation_RSTBnumber} and~\ref{fig:ablation_layernumber}, respectively. It is observed that the PSNR is positively correlated with these three hyper-parameters. For channel number, although the performance keeps increasing, the total number of parameters grows quadratically. To balance the performance and model size, we choose 180 as the channel number in rest experiments. As for RSTB number and layer number, the performance gain becomes saturated gradually. We choose 6 for both of them to obtain a relatively small model.

\vspace{-0.4cm}
\paragraph{Impact of patch size and training image number; model convergence comparison.}
We compare the proposed SwinIR with a representative CNN-based model RCAN to compare the difference of Transformer-based and CNN-based models. From Fig.~
\ref{fig:ablation_patchsize}, one can see that SwinIR performs better than RCAN on different patch sizes, and the PSNR gain becomes larger when the patch size is larger. Fig.~\ref{fig:ablation_dataset} shows the impact of the number of training images. Extra images from Flickr2K are used in training when the percentage is larger than 100\% (800 images). There are two observations. First, as expected, the performance of SwinIR rises with the training image number. Second, different from the observation in IPT that Transformer-based models are heavily relied on large amount of training data, SwinIR achieves better results than CNN-based models using the same training data, even when the dataset is small (\ie, 25\%, 200 images).  We also plot the PSNR during training for both SwinIR and RCAN in Fig.~\ref{fig:ablation_iteration}. It is clear that SwinIR converges faster and better than RCAN, which is contradictory to previous observations that Transformer-based models often suffer from slow model convergence.

\vspace{-0.4cm}
\paragraph{Impact of residual connection and convolution layer in RSTB.}
Table~\ref{tab:ablation_RSTB_design} shows four residual connection variants in RSTB: no residual connection, using $1\times 1$ convolution layer, using $3\times 3$ convolution layer and using three $3\times 3$ convolution layers (channel number of the intermediate layer is set to one fourth of network channel number). From the table, we can have following observations. First, the residual connection in RSTB is important as it improves the PSNR by 0.16dB. Second, using $1\times 1$ convolution brings little improvement maybe because it cannot extract local neighbouring information as $3\times 3$ convolution does. Third, although using three $3\times 3$ convolution layers can reduce the number of parameters, the performance drops slightly.

\begin{table}[t]
\captionsetup{font=small}%
\scriptsize
\center
\begin{center}
\caption{Ablation study on RSTB design.}
\vspace{-2mm}
\label{tab:ablation_RSTB_design}
\begin{tabular}{|l|c|c|c|c|c|}
\hline
Design & No residual & $1\times 1$ conv & $3\times 3$ conv & Three $3\times 3$ conv
\\
\hline
PSNR & 39.42 & 39.45 & 39.58 & 39.56\\
\hline
\end{tabular}
\end{center}
\vspace{-0.3cm}
\end{table}

\subsection{Results on Image SR}

\begin{table*}[t]\scriptsize
\center
\begin{center}
\caption{Quantitative comparison (average PSNR/SSIM) with state-of-the-art methods for \textbf{\underline{classical image SR}} on benchmark datasets. Best and second best performance are in \R{red} and \B{blue} colors, respectively. Results on $\times 8$ are provided in supplementary.}%
\vspace{-2mm}
\label{tab:sr_results}
\begin{tabular}{|l|c|c|c|c|c|c|c|c|c|c|c|c|}
\hline
\multirow{2}{*}{Method} & \multirow{2}{*}{Scale} & \multirow{2}{*}{\makecell{Training\\Dataset}} &  \multicolumn{2}{c|}{Set5~\cite{Set5}} &  \multicolumn{2}{c|}{Set14~\cite{Set14}} &  \multicolumn{2}{c|}{BSD100~\cite{BSD100}} &  \multicolumn{2}{c|}{Urban100~\cite{Urban100}} &  \multicolumn{2}{c|}{Manga109~\cite{Manga109}}  
\\
\cline{4-13}
&  &  & PSNR & SSIM & PSNR & SSIM & PSNR & SSIM & PSNR & SSIM & PSNR & SSIM 
\\
\hline
\hline

RCAN~\cite{zhang2018rcan} & $\times$2 & DIV2K %
& {38.27}
& {0.9614}
& {34.12}
& {0.9216}
& {32.41}
& {0.9027}
& {33.34}
& {0.9384}
& {39.44}
& {0.9786}
\\  
SAN~\cite{dai2019SAN} & $\times$2 & DIV2K %
& {38.31}
& {0.9620}
& {34.07}
& {0.9213}
& {32.42}
& {0.9028}
& {33.10}
& {0.9370}
& {39.32}
& {0.9792}
\\
IGNN~\cite{zhou2020IGNN} & $\times$2 & DIV2K %
& {38.24}
& {0.9613}
& {34.07}
& {0.9217}
& {32.41}
& {0.9025}
& {33.23}
& {0.9383}
& {39.35}
& {0.9786}
\\
HAN~\cite{niu2020HAN} & $\times$2 & DIV2K %
& {38.27}
& {0.9614}
& \B{34.16}
& {0.9217}
& {32.41}
& {0.9027}
& {33.35}
& {0.9385}
& {39.46}
& {0.9785}  
\\ 
NLSA~\cite{mei2021NLSA} & $\times$2 & DIV2K %
& 38.34 
& 0.9618 
& 34.08 
& \B{0.9231}
& 32.43 
& 0.9027 
& \B{33.42}
& \B{0.9394}
& \B{39.59}
& 0.9789
\\
\textbf{\algname{}} (Ours)  & $\times$2  & DIV2K
& \B{38.35}
& \B{0.9620}
& {34.14}
& {0.9227}
& \B{32.44}
& \B{0.9030}
& {33.40}
& {0.9393}
& {39.60}
& \B{0.9792}
\\
\textbf{\algname{}+} (Ours)  & $\times$2  & DIV2K
& \R{38.38}
& \R{0.9621}
& \R{34.24}
& \R{0.9233}
& \R{32.47}
& \R{0.9032}
& \R{33.51}
& \R{0.9401}
& \R{39.70}
& \R{0.9794}
\\
\hdashline
DBPN~\cite{haris2018DBPN} & $\times$2 & DIV2K+Flickr2K%
& 38.09
& 0.9600
& 33.85
& 0.9190
& 32.27
& 0.9000
& 32.55
& 0.9324
& 38.89
& 0.9775        
\\
IPT~\cite{chen2021IPT} & $\times$2  & ImageNet%
& {38.37}
& {-}
& {34.43}
& {-}
& {32.48}
& {-}
& {33.76}
& {-}
& {-}
& {-}
\\
\textbf{\algname{}} (Ours)  & $\times$2  & DIV2K+Flickr2K
& \B{38.42}
& \B{0.9623}
& \B{34.46}
& \B{0.9250}
& \B{32.53}
& \B{0.9041}
& \B{33.81}
& \B{0.9427}
& \B{39.92}
& \B{0.9797}
\\
\textbf{\algname{}+} (Ours)\qquad\qquad\qquad  & $\times$2  & DIV2K+Flickr2K
& \R{38.46}
& \R{0.9624}
& \B{34.61}
& \R{0.9260}
& \R{32.55}
& \R{0.9043}
& \R{33.95}
& \R{0.9433}
& \R{40.02}
& \R{0.9800}
\\

\hline                 
\hline

RCAN~\cite{zhang2018rcan}& $\times$3   & DIV2K
& {34.74}
&{0.9299}
& {30.65}
& {0.8482}
& {29.32}
& {0.8111}
& {29.09}
& {0.8702}
& {34.44}
&{0.9499}
\\
SAN~\cite{dai2019SAN} & $\times$3   & DIV2K
& {34.75}
& {0.9300}
& {30.59}
& {0.8476}
& {29.33}
& {0.8112}
& {28.93}
& {0.8671}
& {34.30}
& {0.9494}
\\
IGNN~\cite{zhou2020IGNN} & $\times$3  & DIV2K
& {34.72}
& {0.9298}
& {30.66}
& {0.8484}
& {29.31}
& {0.8105}
& {29.03}
& {0.8696}
& {34.39}
& {0.9496}
\\

HAN~\cite{niu2020HAN}  & $\times$3   & DIV2K
& {34.75}
& {0.9299}
& {30.67}
& {0.8483}
& {29.32}
& {0.8110}
& {29.10}
& {0.8705}
& {34.48}
& {0.9500}
\\
NLSA~\cite{mei2021NLSA} & $\times$3  & DIV2K
& 34.85 
& 0.9306 
& 30.70 
& 0.8485 
& 29.34 
& 0.8117 
& {29.25}
& {0.8726}
& 34.57 
& 0.9508
\\
\textbf{\algname{}} (Ours)  & $\times$3  & DIV2K
& \B{34.89}
& \B{0.9312}
& \B{30.77}
& \B{0.8503}
& \B{29.37}
& \B{0.8124}
& \B{29.29}
& \B{0.8744}
& \B{34.74}
& \B{0.9518}
\\
\textbf{\algname{}+} (Ours)  & $\times$3  & DIV2K
& \R{34.95}
& \R{0.9316}
& \R{30.83}
& \R{0.8511}
& \R{29.41}
& \R{0.8130}
& \R{29.42}
& \R{0.8761}
& \R{34.92}
& \R{0.9526}
\\
\hdashline
IPT~\cite{chen2021IPT} & $\times$3  & ImageNet
& {34.81}
& {-}
& {30.85}
& {-}
& {29.38}
& {-}
& {29.49}
& {-}
& {-}
& {-}
\\
\textbf{\algname{}} (Ours)  & $\times$3  & DIV2K+Flickr2K
& \B{34.97}
& \B{0.9318}
& \B{30.93}
& \B{0.8534}
& \B{29.46}
& \B{0.8145}
& \B{29.75}
& \B{0.8826}
& \B{35.12}
& \B{0.9537}
\\
\textbf{\algname{}+} (Ours)  & $\times$3  & DIV2K+Flickr2K
& \R{35.04}
& \R{0.9322}
& \R{31.00}
& \R{0.8542}
& \R{29.49}
& \R{0.8150}
& \R{29.90}
& \R{0.8841}
& \R{35.28}
& \R{0.9543}
\\
\hline
\hline

RCAN~\cite{zhang2018rcan}& $\times$4  & DIV2K
& {32.63}
& {0.9002}
& {28.87}
&{0.7889}
& {27.77}
& {0.7436}
&{26.82}
& {0.8087}
&{31.22}
& {0.9173}
\\ 
SAN~\cite{dai2019SAN} & $\times$4  & DIV2K
& {32.64}
& {0.9003}
& {28.92}
& {0.7888}
& {27.78}
& {0.7436}
& {26.79}
& {0.8068}
& {31.18}
& {0.9169}
\\
IGNN~\cite{zhou2020IGNN}  & $\times$4  & DIV2K
& {32.57}
& {0.8998}
& {28.85}
& {0.7891}
& {27.77}
& {0.7434}
& {26.84}
& {0.8090}
& {31.28}
& {0.9182}
\\

HAN~\cite{niu2020HAN}  & $\times$4  & DIV2K
& {32.64}
& {0.9002}
& {28.90}
& {0.7890}
& {27.80}
& {0.7442}
& {26.85}
& {0.8094}
& {31.42}
& {0.9177}
\\
NLSA~\cite{mei2021NLSA} & $\times$4 & DIV2K
& 32.59 
& 0.9000 
& 28.87 
& 0.7891 
& 27.78 
& 0.7444 
& {26.96}
& {0.8109}
& 31.27 
& 0.9184
\\
\textbf{\algname{}} (Ours)  & $\times$4  & DIV2K
& \B{32.72}
& \B{0.9021}
& \B{28.94}
& \B{0.7914}
& \B{27.83}
& \B{0.7459}
& \B{27.07}
& \B{0.8164}
& \B{31.67}
& \B{0.9226}
\\
\textbf{\algname{}+} (Ours)  & $\times$4  & DIV2K
& \R{32.81}
& \R{0.9029}
& \R{29.02}
& \R{0.7928}
& \R{27.87}
& \R{0.7466}
& \B{27.21}
& \R{0.8187}
& \R{31.88}
& \R{0.9423}
\\
\hdashline
DBPN~\cite{haris2018DBPN} & $\times$4 & DIV2K+Flickr2K
& 32.47
& 0.8980
& 28.82
& 0.7860
& 27.72
& 0.7400
& 26.38
& 0.7946
& 30.91
& 0.9137
\\
IPT~\cite{chen2021IPT} & $\times$4 & ImageNet
& {32.64}
& {-}
& {29.01}
& {-}
& {27.82}
& {-}
& {27.26}
& {-}
& {-}
& {-}
\\
RRDB~\cite{wang2018esrgan} & $\times$4 & DIV2K+Flickr2K
& {32.73}
& {0.9011 }
& {28.99}
& {0.7917}
& {27.85}
& {0.7455}
& {27.03}
& {0.8153}
& {31.66}
& {0.9196}
\\
\textbf{\algname{}} (Ours)  & $\times$4  & DIV2K+Flickr2K
& \B{32.92}
& \B{0.9044}
& \B{29.09}
& \B{0.7950}
& \B{27.92}
& \B{0.7489}
& \B{27.45}
& \B{0.8254}
& \B{32.03}
& \B{0.9260}
\\
\textbf{\algname{}+} (Ours)  & $\times$4  & DIV2K+Flickr2K
& \R{32.93}
& \R{0.9043}
& \R{29.15}
& \R{0.7958}
& \R{27.95}
& \R{0.7494}
& \R{27.56}
& \R{0.8273}
& \R{32.22}
& \R{0.9273}
\\
\hline             
\end{tabular}
\end{center}
\vspace{1mm}
\end{table*}

\begin{figure*}[!htbp]
	\captionsetup{font=small}
	
	\centering
	\scriptsize
	
	\renewcommand{\h}{0.105}
	\renewcommand{\wa}{0.12}
	\newcommand{\wb}{0.16}
	\renewcommand{\g}{-0.7mm}
	\renewcommand{\tabcolsep}{1.8pt}
	\renewcommand{\arraystretch}{1}
	\resizebox{1.00\linewidth}{!} {
		\begin{tabular}{cc}
			
			\renewcommand{\name}{figures/sr/Urban_012/img_012_}
			\renewcommand{\h}{0.078}
			\renewcommand{\w}{0.176}
			\begin{tabular}{cc}
				\begin{adjustbox}{valign=t}
					\begin{tabular}{c}
						\includegraphics[height=0.185\textwidth, width=0.374\textwidth]{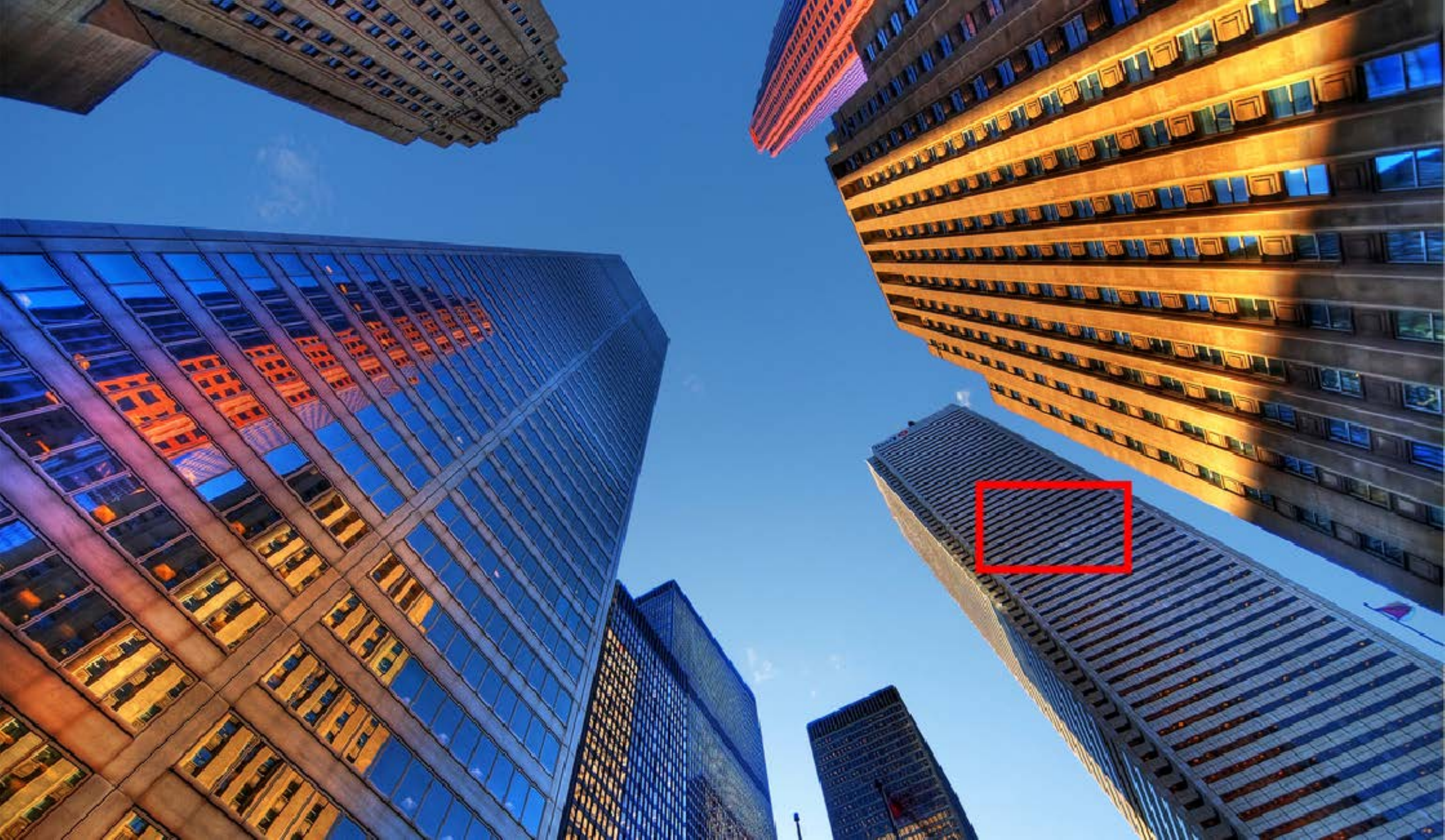}
						\\
						Urban100 ($4\times$):img\_012
					\end{tabular}
				\end{adjustbox}
				\begin{adjustbox}{valign=t}
					\begin{tabular}{cccccc}
						\includegraphics[height=\h \textwidth, width=\w \textwidth]{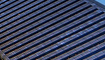} \hspace{\g} &
						\includegraphics[height=\h \textwidth, width=\w \textwidth]{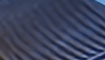} \hspace{\g} &
						\includegraphics[height=\h \textwidth, width=\w \textwidth]{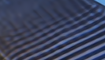} \hspace{\g} &
						\includegraphics[height=\h \textwidth, width=\w \textwidth]{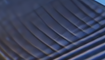} \hspace{\g} &
						\includegraphics[height=\h \textwidth, width=\w \textwidth]{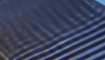} 
						\\
						HR \hspace{\g} &
						VDSR~\cite{kim2016vdsr} \hspace{\g} &
						EDSR~\cite{lim2017edsr} \hspace{\g} &
						RDN~\cite{zhang2018RDN} & OISR~\cite{he2019ode} \hspace{\g}
						\\
						\vspace{-1.5mm}
						\\
						
						\includegraphics[height=\h \textwidth, width=\w \textwidth]{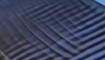} \hspace{\g} &
						\includegraphics[height=\h \textwidth, width=\w \textwidth]{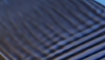} \hspace{\g} &
						\includegraphics[height=\h \textwidth, width=\w \textwidth]{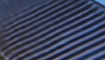}
						\hspace{\g} &		
						\includegraphics[height=\h \textwidth, width=\w \textwidth]{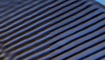} \hspace{\g} &
						\includegraphics[height=\h \textwidth, width=\w \textwidth]{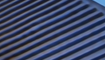} 
						\\ 
						
						SAN~\cite{dai2019SAN} \hspace{\g} &
						RNAN~\cite{zhang2019RNAN}  \hspace{\g} &
						IGNN~\cite{zhou2020IGNN} \hspace{\g} &
						IPT~\cite{chen2021IPT}
						& \textbf{SwinIR} (ours)
						\\
					\end{tabular}
				\end{adjustbox}
			\end{tabular}
			
		\end{tabular}
	}\vspace{-2mm}
	\caption{Visual comparison of \textbf{\underline{bicubic image SR}} ($\times 4$) methods. Compared images are derived from~\cite{chen2021IPT}. Best viewed by zooming.}
	\label{fig:sr_visual}
\end{figure*}

\begin{table*}[!t]\scriptsize
\center
\begin{center}
\caption{Quantitative comparison (average PSNR/SSIM) with state-of-the-art methods for \textbf{\underline{lightweight image SR}} on benchmark datasets. Best and second best performance are in \R{red} and \B{blue} colors, respectively.}
\vspace{-2mm}
\label{tab:lightweight_sr_results}
\begin{tabular}{|l|c|c|c|c|c|c|c|c|c|c|c|c|c|}
\hline
\multirow{2}{*}{Method} & \multirow{2}{*}{Scale} & \multirow{2}{*}{\#Params} & \multirow{2}{*}{\#Mult-Adds} &  \multicolumn{2}{c|}{Set5~\cite{Set5}} &  \multicolumn{2}{c|}{Set14~\cite{Set14}} &  \multicolumn{2}{c|}{BSD100~\cite{BSD100}} &  \multicolumn{2}{c|}{Urban100~\cite{Urban100}} &  \multicolumn{2}{c|}{Manga109~\cite{Manga109}}  
\\
\cline{5-14}
&  &  &  & PSNR & SSIM & PSNR & SSIM & PSNR & SSIM & PSNR & SSIM & PSNR & SSIM 
\\
\hline
\hline
CARN~\cite{ahn2018CARN} & $\times$2  & 1,592K & 222.8G
& 37.76
& 0.9590
& 33.52
& 0.9166
& 32.09
& 0.8978
& 31.92
& 0.9256
& 38.36
& 0.9765
\\
FALSR-A~\cite{chu2021fast} & $\times$2  & 1,021K & 234.7G
& 37.82
& 0.959
& 33.55
& 0.9168
& 32.1
& 0.8987 
& 31.93 
& 0.9256
& -
& -
\\
IMDN~\cite{hui2019imdn} & $\times$2  & 694K & 158.8G
& 38.00
& 0.9605
& 33.63
& 0.9177
& 32.19
& 0.8996
& 32.17
& 0.9283
& \B{38.88}
& \B{0.9774}
\\
LAPAR-A~\cite{li2021lapar} & $\times$2  & 548K & 171.0G
& 38.01
& 0.9605
& 33.62
& 0.9183
& 32.19
& 0.8999
& 32.10
& 0.9283
& 38.67
& 0.9772
\\
LatticeNet~\cite{luo2020latticenet} & $\times$2  & 756K & 169.5G
& \R{38.15}
& \B{0.9610}
& \B{33.78}
& \B{0.9193}
& \B{32.25}
& \B{0.9005}
& \B{32.43}
& \B{0.9302}
& -
& -
\\
\textbf{\algname{}} (Ours)\qquad\qquad  & $\times$2  & 878K & {195.6G} %
& \B{38.14}
& \R{0.9611}
& \R{33.86}
& \R{0.9206}
& \R{32.31}
& \R{0.9012}
& \R{32.76}
& \R{0.9340}
& \R{39.12}
& \R{0.9783}
\\
\hline   
CARN~\cite{ahn2018CARN} & $\times$3  & 1,592K  & 118.8G
& 34.29
& 0.9255
& 30.29
& 0.8407
& 29.06
& 0.8034
& 28.06
& 0.8493
& 33.50 
& 0.9440
\\ 
IMDN~\cite{hui2019imdn} & $\times$3  & 703K  & 71.5G 
& 34.36
& 0.9270
& 30.32
& 0.8417
& 29.09
& 0.8046
& 28.17
& 0.8519
& \B{33.61}
& \B{0.9445}
\\ 
LAPAR-A~\cite{li2021lapar} & $\times$3  & 544K & 114.0G
& 34.36
& 0.9267
& 30.34
& 0.8421
& 29.11
& 0.8054
& 28.15
& 0.8523
& 33.51
& 0.9441
\\
LatticeNet~\cite{luo2020latticenet} & $\times$3  & 765K & 76.3G 
& \B{34.53}
& \B{0.9281}
& \B{30.39}
& \B{0.8424}
& \B{29.15}
& \B{0.8059}
& \B{28.33}
& \B{0.8538}
& -
& -
\\
\textbf{\algname{}} (Ours)  & $\times$3  & 886K & {87.2G} %
& \R{34.62}
& \R{0.9289}
& \R{30.54}
& \R{0.8463}
& \R{29.20}
& \R{0.8082}
& \R{28.66}
& \R{0.8624}
& \R{33.98}
& \R{0.9478}
\\
\hline 
\hline
CARN~\cite{ahn2018CARN} & $\times$4  & 1,592K & 90.9G
& 32.13
& 0.8937
& 28.60
& 0.7806
& 27.58
& 0.7349
& 26.07 
& 0.7837
& \B{30.47}
& \B{0.9084}
\\
IMDN~\cite{hui2019imdn} & $\times$4  & 715K & 40.9G
& 32.21
& 0.8948
& 28.58
& 0.7811
& 27.56
& 0.7353 
& 26.04
& 0.7838
& 30.45
& 0.9075
\\
LAPAR-A~\cite{li2021lapar} & $\times$4  & 659K & 94.0G
& 32.15
& 0.8944
& 28.61
& 0.7818
& 27.61
& 0.7366
& 26.14
& 0.7871
& 30.42
& 0.9074
\\
LatticeNet~\cite{luo2020latticenet} & $\times$4  & 777K & 43.6G
& \B{32.30}
& \B{0.8962}
& \B{28.68}
& \B{0.7830}
& \B{27.62}
& \B{0.7367}
& \B{26.25}
& \B{0.7873}
& -
& -
\\
\textbf{\algname{}} (Ours)  & $\times$4  & 897K & {49.6G} %
& \R{32.44}
& \R{0.8976}
& \R{28.77}
& \R{0.7858}
& \R{27.69}
& \R{0.7406}
& \R{26.47}
& \R{0.7980}
& \R{30.92}
& \R{0.9151}
\\
\hline 
\end{tabular}
\end{center}
\end{table*}

\begin{figure*}[!tbp]\footnotesize
\captionsetup{font=small}
\hspace{-0.20cm}
\begin{tabular}{c@{\extracolsep{0em}}c@{\extracolsep{0.05em}}c@{\extracolsep{0.05em}}c@{\extracolsep{0.05em}}c@{\extracolsep{0.05em}}@{\extracolsep{0.05em}}c@{\extracolsep{0.05em}}c}

        \includegraphics[width=0.158\textwidth]{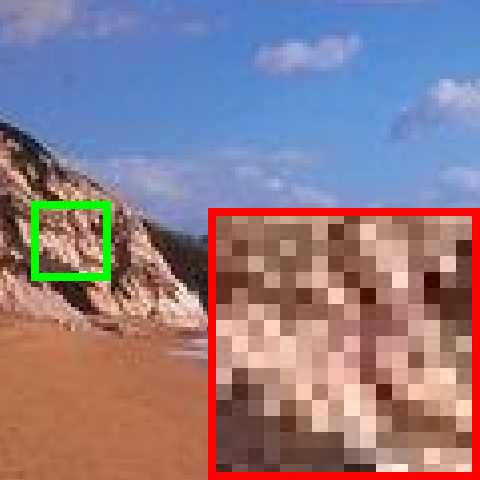}~
		&\includegraphics[width=0.158\textwidth]{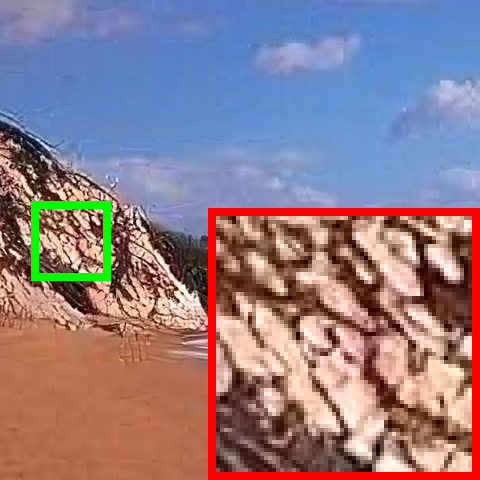}~
        &\includegraphics[width=0.158\textwidth]{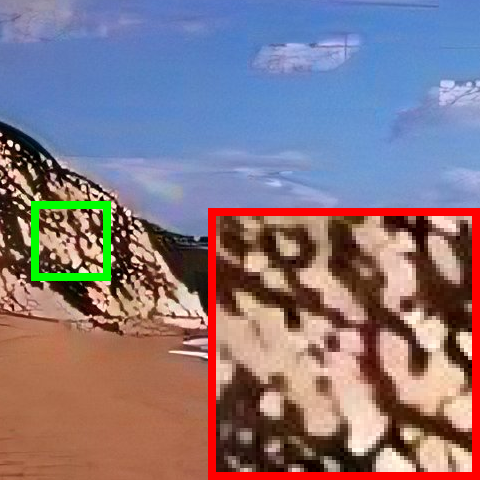}~
		&\includegraphics[width=0.158\textwidth]{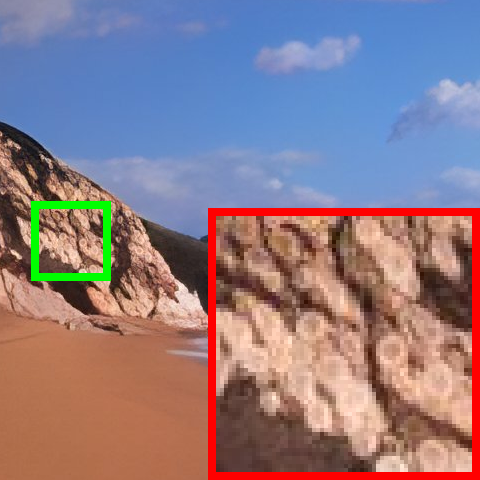}~
		&\includegraphics[width=0.158\textwidth]{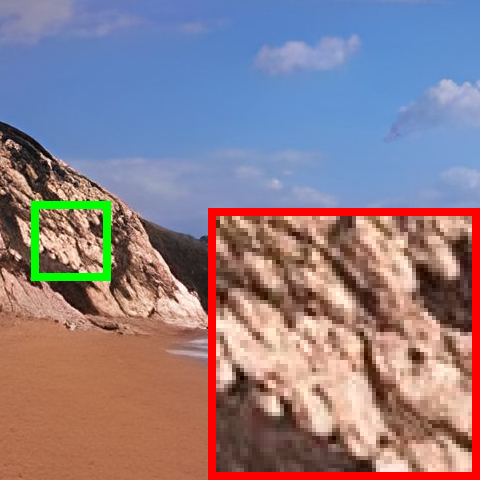}~
		&\includegraphics[width=0.158\textwidth]{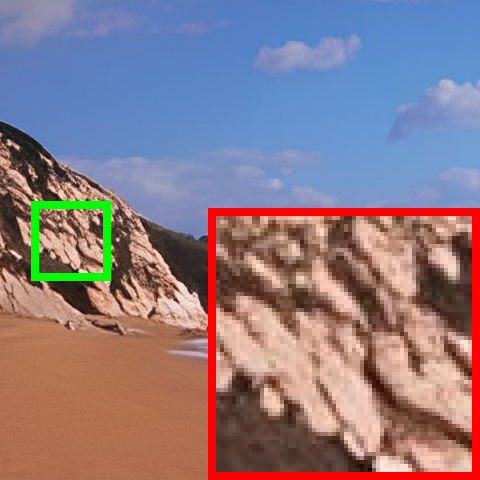}
		\\
		
        \includegraphics[width=0.158\textwidth]{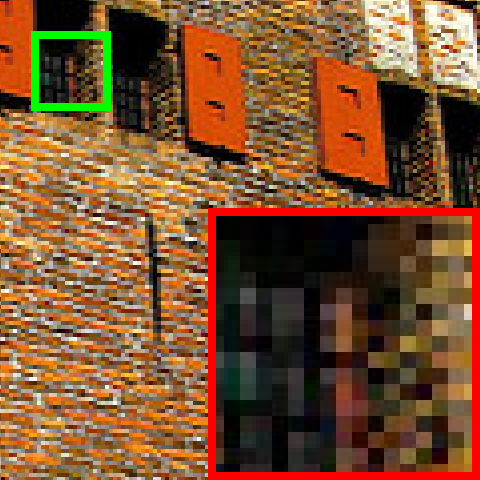}~
		&\includegraphics[width=0.158\textwidth]{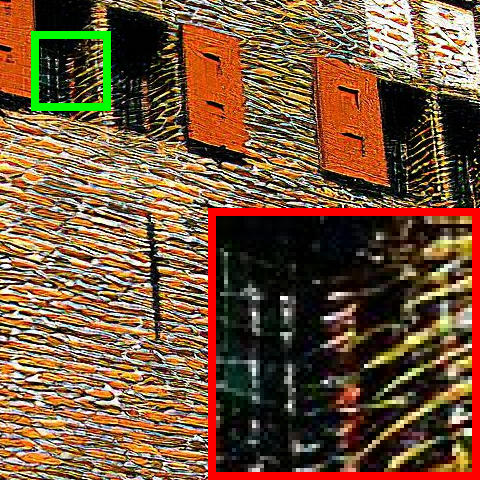}~
        &\includegraphics[width=0.158\textwidth]{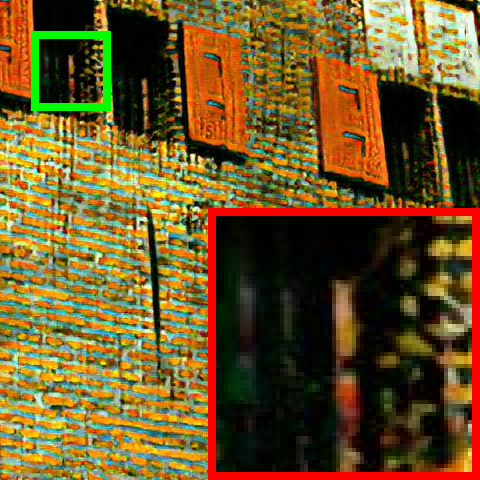}~
		&\includegraphics[width=0.158\textwidth]{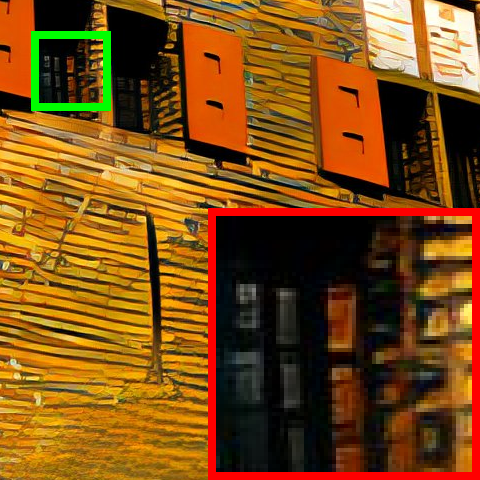}~
		&\includegraphics[width=0.158\textwidth]{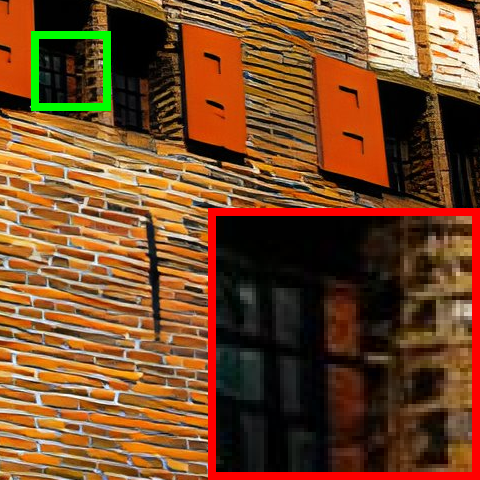}~
		&\includegraphics[width=0.158\textwidth]{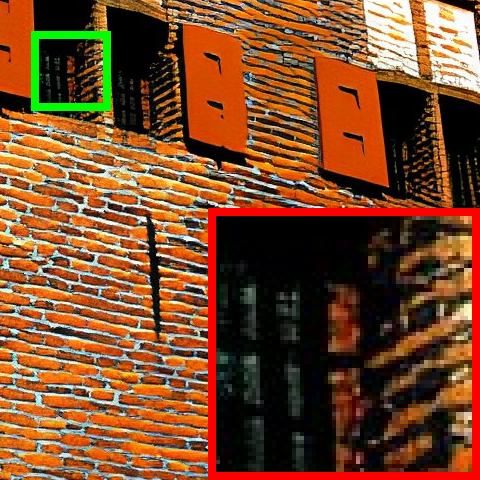}
		\\
 LR ($\times$4) & ESRGAN~\cite{wang2018esrgan}  & RealSR~\cite{ji2020realsr} &   BSRGAN~\cite{kai2021bsrgan} & Real-ESRGAN~\cite{wang2021realESRGAN} & \textbf{SwinIR} (ours)\\
	\end{tabular}
    \vspace{-0.2cm}
	\caption{Visual comparison of \textbf{\underline{real-world image SR}} ($\times 4$) methods on real-world images.%
	}
	\label{fig:reasr_visual}
	\vspace{-0.1cm}
\end{figure*}

\begin{table*}[h]\scriptsize
\center
\begin{center}
\caption{Quantitative comparison (average PSNR/SSIM/PSNR-B) with state-of-the-art methods for \textbf{\underline{JPEG compression artifact reduction}} on benchmark datasets. Best and second best performance are in \R{red} and \B{blue} colors, respectively.}
\vspace{-2mm}
\label{tab:car_results}
\begin{tabular}
{|p{0.7cm}<{\centering}|p{0.2cm}<{\centering}|p{1.85cm}<{\centering}|p{1.85cm}<{\centering}|p{1.85cm}<{\centering}|p{1.85cm}<{\centering}|p{1.45cm}<{\centering}|p{1.85cm}<{\centering}|p{1.85cm}<{\centering}|}
\hline
Dataset & $q$ 
& ARCNN~\cite{dong2015compression} 
& DnCNN-3~\cite{zhang2017DnCNN} 
& QGAC~\cite{ehrlich2020quantization}
& RNAN~\cite{zhang2019RNAN}
& RDN~\cite{zhang2020RDNIR}
& DRUNet~\cite{zhang2021DPIR}
& \textbf{SwinIR} (ours)\\
\hline
\hline
\multirow{4}{*}{\makecell{Classic5\\\cite{foi2007Classic5}}} & 10
& 29.03/0.7929/28.76
& 29.40/0.8026/29.13
& 29.84/0.8370/29.43
& 29.96/0.8178/29.62
& 30.00/0.8188/-
& \B{30.16/0.8234/29.81}
& \R{30.27/0.8249/29.95}
\\
& 20
& 31.15/0.8517/30.59
& 31.63/0.8610/31.19
& 31.98/0.8850/31.37
& 32.11/0.8693/31.57
& 32.15/0.8699/-
& \B{32.39/0.8734/31.80}
& \R{32.52/0.8748/31.99}
\\
& 30
& 32.51/0.8806/31.98
& 32.91/0.8861/32.38
& 33.22/0.9070/32.42
& 33.38/0.8924/32.68
& 33.43/0.8930/-
& \B{33.59/0.8949/32.82}
& \R{33.73/0.8961/33.03}
\\
& 40
& 33.32/0.8953/32.79
& 33.77/0.9003/33.20
& -
& 34.27/0.9061/33.4
& 34.27/0.9061/-
& \B{34.41/0.9075/33.51}
& \R{34.52/0.9082/33.66}
\\
\hline
\multirow{4}{*}{\makecell{LIVE1\\\cite{sheikh2005live}}} & 10
& 28.96/0.8076/28.77
& 29.19/0.8123/28.90
& 29.53/0.8400/29.15 
& 29.63/0.8239/29.25
& 29.67/0.8247/-
& \B{29.79/0.8278/29.48}
& \R{29.86/0.8287/29.50}
\\
& 20
& 31.29/0.8733/30.79
& 31.59/0.8802/31.07
& 31.86/0.9010/31.27
& 32.03/0.8877/31.44
& 32.07/0.8882/-
& \B{32.17/0.8899/31.69}
& \R{32.25/0.8909/31.70}
\\
& 30
& 32.67/0.9043/32.22
& 32.98/0.9090/32.34
& 33.23/0.9250/32.50
& 33.45/0.9149/32.71
& 33.51/0.9153/-
& \B{33.59/0.9166/32.99}
& \R{33.69/0.9174/33.01}
\\
& 40
& 33.63/0.9198/33.14
& 33.96/0.9247/33.28
& -
& 34.47/0.9299/33.66
& 34.51/0.9302/-
& \B{34.58/0.9312/}\R{33.93}
& \R{34.67/0.9317/}\B{33.88}
\\
\hline             
\end{tabular}
\end{center}
\vspace{-2mm}
\end{table*}

\vspace{-2mm}
\paragraph{Classical image SR.}
Table~\ref{tab:sr_results} shows the quantitative comparisons between SwinIR (middle size) and state-of-the-art methods: DBPN~\cite{haris2018DBPN}, RCAN~\cite{zhang2018rcan}, RRDB~\cite{wang2018esrgan}, SAN~\cite{dai2019SAN}, IGNN~\cite{zhou2020IGNN}, HAN~\cite{niu2020HAN}, NLSA~\cite{mei2021NLSA} and IPT~\cite{chen2021IPT}. As one can see, when trained on DIV2K, SwinIR achieves best performance on almost all five benchmark datasets for all scale factors. The maximum PSNR gain reaches 0.26dB on Manga109 for scale factor 4. Note that RCAN and HAN introduce channel and spatial attention, IGNN proposes adaptive patch feature aggregation, and NLSA is based on the non-local attention mechanism. However, all these CNN-based attention mechanisms perform worse than the proposed Transformer-based SwinIR, which indicates the effectiveness of the proposed model.
When we train SwinIR on a larger dataset (DIV2K+Flickr2K), the performance further increases by a large margin
(up to 0.47dB), achieving better accuracy than the same Transformer-based model IPT, even though IPT utilizes ImageNet (more than 1.3M images) in training and has huge number of parameters (115.5M). In contrast, SwinIR has a small number of parameters (11.8M) even compared with state-of-the-art CNN-based models (15.4$\sim$44.3M). As for runtime, representative CNN-based model RCAN, IPT and SwinIR take about 0.2, 4.5s and 1.1s to test on a $1,024\times 1,024$ image, respectively. %
Visual comparisons are show in Fig.~\ref{fig:sr_visual}. SwinIR can restore high-frequency details and alleviate the blurring artifacts, resulting in sharp and natural edges. In contrast, most CNN-based methods produces blurry images or even incorrect textures. IPT generates better images compared with CNN-based methods, but it suffers from image distortions and border artifact.

\vspace{-4mm}
\paragraph{Lightweight image SR.} We also provide comparison of SwinIR (small size) with state-of-the-art lightweight image SR methods: CARN~\cite{ahn2018CARN}, FALSR-A~\cite{chu2021fast}, IMDN~\cite{hui2019imdn}, LAPAR-A~\cite{li2021lapar} and LatticeNet~\cite{luo2020latticenet}. In addition to PSNR and SSIM, we also report the total numbers of parameters and multiply-accumulate operations (evaluated on a $1280\times 720$ HQ image) to compare the model size and computational complexity of different models. As shown in Table~\ref{tab:lightweight_sr_results}, SwinIR outperforms competitive methods by a PSNR margin of up to 0.53dB on different benchmark datasets, with similar total numbers of parameters and multiply-accumulate operations. This indicates that the SwinIR architecture is highly efficient for image restoration.

\vspace{-4mm}
\paragraph{Real-world image SR.} The ultimate goal of image SR is for real-world applications. Recently, Zhang~\etal~\cite{kai2021bsrgan} proposed a practical degradation model BSRGAN for real-world image SR and achieved surprising results in real scenarios\footnote{\url{https://github.com/cszn/BSRGAN}}. To test the performance of SwinIR for real-world SR, \textbf{we re-train SwinIR by using the same degradation model as BSRGAN} for low-quality image synthesis. Since there is no ground-truth high-quality images, we only provide visual comparison with representative bicubic model ESRGAN~\cite{wang2018esrgan} and state-of-the-art real-world image SR models RealSR~\cite{ji2020realsr}, BSRGAN~\cite{kai2021bsrgan} and Real-ESRGAN~\cite{wang2021realESRGAN}. As shown in Fig.~\ref{fig:reasr_visual}, SwinIR produces visually pleasing images with clear and sharp edges, whereas other compared methods may suffer from unsatisfactory artifacts. In addition, to exploit the full potential of SwinIR for real applications, we further propose a large model and train it on much larger datasets. Experiments show that it can deal with more complex corruptions and achieves even better performance on real-world images than the current model. Due to page limit, the details are given in our project page \url{https://github.com/JingyunLiang/SwinIR}.

\subsection{Results on JPEG Compression Artifact Reduction}
Table~\ref{tab:car_results} shows the comparison of SwinIR with state-of-the-art JPEG compression artifact reduction methods: ARCNN~\cite{dong2015compression}, DnCNN-3~\cite{zhang2017DnCNN}, QGAC~\cite{ehrlich2020quantization}, RNAN~\cite{zhang2019RNAN}, RDN~\cite{zhang2020RDNIR} and DRUNet~\cite{zhang2021DPIR}. All of compared methods are CNN-based models. Following~\cite{zhang2020RDNIR, zhang2021DPIR}, we test different methods on two benchmark datasets (Classic5~\cite{foi2007Classic5} and LIVE1~\cite{sheikh2005live}) for JPEG quality factors 10, 20, 30 and 40. As we can see, the proposed SwinIR has average PSNR gains of at least 0.11dB and 0.07dB on two testing datasets for different quality factors. Besides, compared with the previous best model DRUNet, SwinIR only has 11.5M parameters, while DRUNet is a large model that has 32.7M parameters.

\begin{table*}[!t]\scriptsize
\center
\begin{center}
\caption{Quantitative comparison (average PSNR) with state-of-the-art methods for \textbf{\underline{grayscale image denoising}} on benchmark datasets. Best and second best performance are in \R{red} and \B{blue} colors, respectively.}
\vspace{-2mm}
\label{tab:denoising_grayscale_results}
\begin{tabular}
{|c|c|c|c|c|c|c|c|c|c|c|c|c|c|}
\hline
~Dataset~ & ~$\sigma$~ & \makecell{BM3D\\\cite{dabov2007bm3d}} & \makecell{WNNM\\\cite{gu2014weighted}} & \makecell{DnCNN\\\cite{zhang2017DnCNN}} & 
\makecell{IRCNN\\\cite{zhang2017IRCNN}} &
\makecell{FFDNet\\\cite{zhang2018ffdnet}} &
\makecell{N3Net\\\cite{plotz2018N3Net}} &  \makecell{NLRN\\\cite{liu2018NLRN}} &
\makecell{FOCNet\\\cite{jia2019focnet}} &
\makecell{RNAN\\\cite{zhang2019RNAN}} &  
\makecell{MWCNN\\\cite{liu2018MWCNN}} & 
\makecell{DRUNet\\\cite{zhang2021DPIR}} &
\makecell{\textbf{SwinIR} (ours)}
\\
\hline
\hline
\multirow{3}{*}{\makecell{Set12\\\cite{zhang2017DnCNN}}} & 15
& 32.37 
& 32.70
& 32.86
& 32.76 
& 32.75
& -
& 33.16 
& 33.07
& -
& 33.15
& \B{33.25} %
& \R{33.36}
\\
& 25
& 29.97
& 30.28
& 30.44
& 30.37
& 30.43
& 30.55
& 30.80
& 30.73
& -
& 30.79
& \B{30.94}
& \R{31.01}
\\
& 50
& 26.72
& 27.05
& 27.18
& 27.12
& 27.32 
& 27.43
& 27.64
& 27.68
& 27.70
& 27.74
& \B{27.90}
& \R{27.91}
\\
\hline
\multirow{3}{*}{\makecell{BSD68\\\cite{BSD68}}} & 15
& 31.08
& 31.37
& 31.73
& 31.63
& 31.63
& -
& 31.88
& 31.83
& -
& 31.86
& \B{31.91}
& \R{31.97}
\\
& 25
& 28.57
& 28.83
& 29.23
& 29.15
& 29.19
& 29.30
& 29.41
& 29.38
& -
& 29.41
& \B{29.48}
& \R{29.50}
\\
& 50
& 25.60
& 25.87
& 26.23
& 26.19
& 26.29
& 26.39
& 26.47
& 26.50
& 26.48
& 26.53
& \R{26.59}
& \B{26.58}
\\
\hline
\multirow{3}{*}{\makecell{\hspace{-0.15cm}Urban100\\\cite{Urban100}}} & 15
& 32.35
& 32.97
& 32.64
& 32.46
& 32.40
& -
& 33.45
& 33.15
& -
& 33.17
& \B{33.44}
& \R{33.70}
\\
& 25
& 29.70
& 30.39
& 29.95
& 29.80
& 29.90
& 30.19
& 30.94
& 30.64
& -
& 30.66
& \B{31.11}
& \R{31.30}
\\
& 50
& 25.95
& 26.83
& 26.26
& 26.22
& 26.50
& 26.82
& 27.49
& 27.40
& 27.65
& 27.42
& \B{27.96}
& \R{27.98}
\\
\hline             
\end{tabular}
\end{center}
\end{table*}

\begin{table*}[!t]\scriptsize
\center
\begin{center}
\caption{Quantitative comparison (average PSNR) with state-of-the-art methods for \textbf{\underline{color image denoising}} on benchmark datasets. Best and second best performance are in \R{red} and \B{blue} colors, respectively.}
\vspace{-2mm}
\label{tab:denoising_color_results}
\begin{tabular}{|c|c|c|c|c|c|c|c|c|c|c|c|c|c|}
\hline
~~~~~Dataset~~~~~ & ~$\sigma$~ & \makecell{BM3D\\\cite{dabov2007bm3d}} & \makecell{DnCNN\\\cite{zhang2017DnCNN}} &
\makecell{IRCNN\\\cite{zhang2017IRCNN}} &
\makecell{FFDNet\\\cite{zhang2018ffdnet}} &
\makecell{DSNet\\\cite{peng2019dsnet}} & 
\makecell{RPCNN\\\cite{xia2020rpcnn}} & 
\makecell{BRDNet\\\cite{tian2020BRDnet}} & 
\makecell{RNAN\\\cite{zhang2019RNAN}} &
\makecell{RDN\\\cite{zhang2020RDNIR}} &
\makecell{IPT\\\cite{chen2021IPT}} &
\makecell{DRUNet\\\cite{zhang2021DPIR}} &
\makecell{\textbf{SwinIR} (ours)}
\\
\hline
\hline
\multirow{3}{*}{\makecell{CBSD68\\\cite{BSD68}}} & 15
& 33.52   
& 33.90
& 33.86
& 33.87
& 33.91
& -
& 34.10
& -
& -
& -
& \B{34.30}
& \R{34.42}
\\
& 25
& 30.71  
& 31.24 
& 31.16
& 31.21
& 31.28
& 31.24
& 31.43
& -
& -
& -
& \B{31.69}
& \R{31.78}
\\
& 50
& 27.38   
& 27.95
& 27.86
& 27.96
& 28.05
& 28.06
& 28.16
& 28.27
& 28.31
& 28.39
& \B{28.51}
& \R{28.56}
\\
\hline
\multirow{3}{*}{\makecell{Kodak24\\\cite{franzen1999kodak}}} & 15
& 34.28  
& 34.60 
& 34.69
& 34.63
& 34.63
& -
& 34.88
& -
& -
& -
& \B{35.31}
& \R{35.34}
\\
& 25
& 32.15   
& 32.14
& 32.18
& 32.13
& 32.16
& 32.34
& 32.41
& -
& -
& -
& \B{32.89}
& \R{32.89}
\\
& 50
& 28.46   
& 28.95
& 28.93
& 28.98
& 29.05
& 29.25
& 29.22
& 29.58
& 29.66
& 29.64
& \R{29.86}
& \B{29.79}
\\
\hline
\multirow{3}{*}{\makecell{McMaster\\\cite{zhang2011McMaster}}} & 15
& 34.06   
& 33.45
& 34.58
& 34.66
& 34.67
& -
& 35.08
& -
& -
& -
& \B{35.40}
& \R{35.61}
\\
& 25
& 31.66  
& 31.52
& 32.18
& 32.35
& 32.40
& 32.33
& 32.75
& -
& -
& -
& \B{33.14}
& \R{33.20}
\\
& 50
& 28.51  
& 28.62 
& 28.91
& 29.18
& 29.28
& 29.33
& 29.52
& 29.72
& -
& 29.98
& \B{30.08}
& \R{30.22}
\\
\hline
\multirow{3}{*}{\makecell{Urban100\\\cite{Urban100}}} & 15
& 33.93
& 32.98
& 33.78
& 33.83
& -
& -
& 34.42
& -
& -
& -
& \B{34.81}
& \R{35.13}
\\
& 25
& 31.36
& 30.81
& 31.20
& 31.40
& -
& 31.81
& 31.99
& -
& -
& -
& \B{32.60}
& \R{32.90}
\\
& 50
& 27.93
& 27.59
& 27.70
& 28.05
& -
& 28.62
& 28.56
& 29.08
& 29.38
& \B{29.71} %
& {29.61}
& \R{29.82}
\\
\hline             
\end{tabular}
\end{center}
\end{table*}

\begin{figure*}[!htbp]
\captionsetup{font=small}
\scriptsize
\hspace{-0.2cm}
\begin{tabular}{c@{\extracolsep{0em}}@{\extracolsep{0.05em}}c@{\extracolsep{0.05em}}c@{\extracolsep{0.05em}}c@{\extracolsep{0.05em}}c@{\extracolsep{0.00em}}c@{\extracolsep{0.00em}}}
		\includegraphics[width=0.157\textwidth]{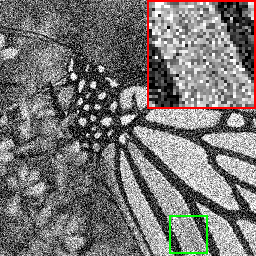}~~
		&\includegraphics[width=0.157\textwidth]{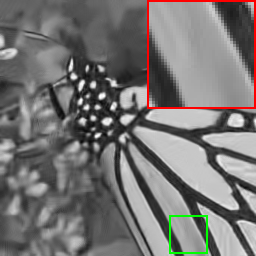}~~
		&\includegraphics[width=0.157\textwidth]{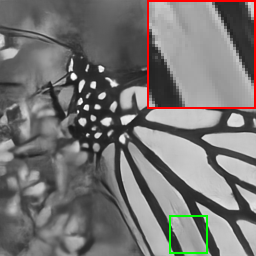}~~
		&\includegraphics[width=0.157\textwidth]{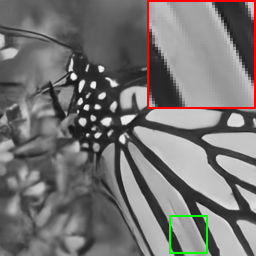}~~
		&\includegraphics[width=0.157\textwidth]{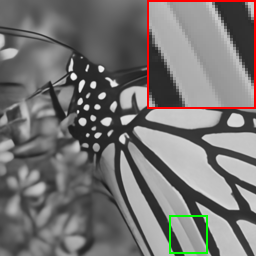}~~
		&\includegraphics[width=0.157\textwidth]{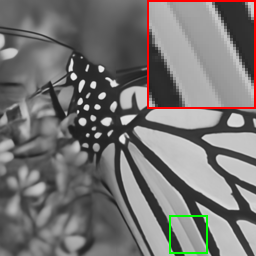}~~\\

 Noisy & BM3D~\cite{dabov2007bm3d} & DnCNN~\cite{zhang2017DnCNN} & FFDNet~\cite{zhang2018ffdnet}  & DRUNet~\cite{zhang2021DPIR}  & ~\textbf{SwinIR} (ours) \\
\end{tabular}
\vspace{-0.2cm}
\caption{Visual comparison of \textbf{\underline{grayscale image denoising}} (noise level 50) methods on image ``\emph{Monarch}'' from Set12~\cite{zhang2017DnCNN}. Compared images are derived from~\cite{zhang2021DPIR}.}\label{fig:denoising_gray_result}
\end{figure*}

\begin{figure*}[!htbp]
\captionsetup{font=small}
\scriptsize
\hspace{-0.2cm}
\begin{tabular}{c@{\extracolsep{0em}}@{\extracolsep{0.05em}}c@{\extracolsep{0.05em}}c@{\extracolsep{0.05em}}c@{\extracolsep{0.05em}}c@{\extracolsep{0.00em}}c@{\extracolsep{0.00em}}}
		\includegraphics[width=0.157\textwidth]{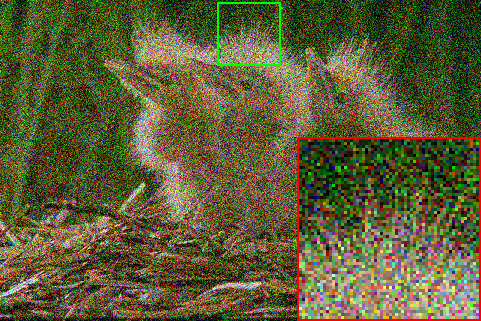}~~
		&\includegraphics[width=0.157\textwidth]{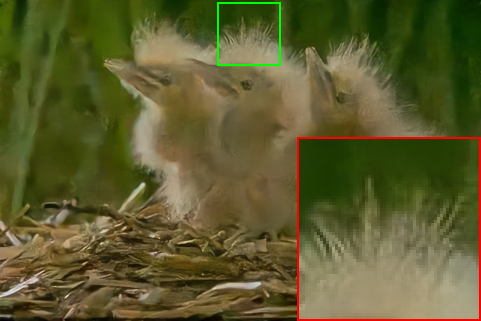}~~
		&\includegraphics[width=0.157\textwidth]{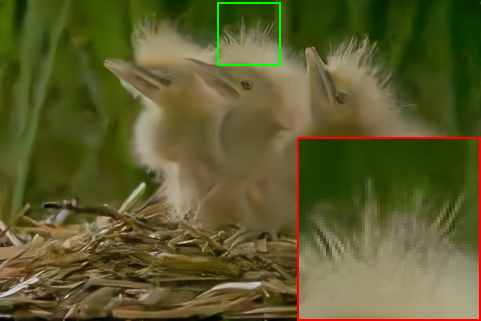}~~
		&\includegraphics[width=0.157\textwidth]{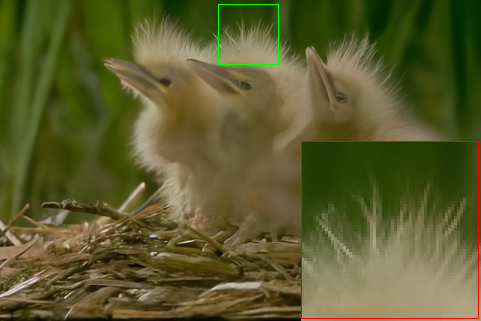}~~
		&\includegraphics[width=0.157\textwidth]{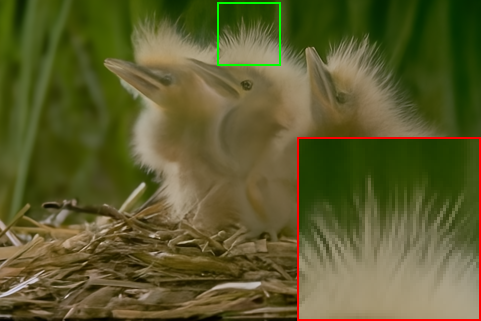}~~
		&\includegraphics[width=0.157\textwidth]{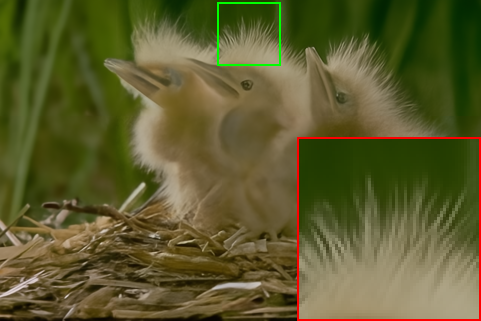}~~\\

 Noisy & DnCNN~\cite{zhang2017DnCNN} & FFDNet~\cite{zhang2018ffdnet}  & IPT~\cite{chen2021IPT} & DRUNet~\cite{zhang2021DPIR}  & ~\textbf{SwinIR} (ours) \\
\end{tabular}
\vspace{-0.2cm}
\caption{Visual comparison of \textbf{\underline{color image denoising}} (noise level 50) methods on image ``\emph{163085}'' from CBSD68~\cite{BSD68}. Compared images are derived from~\cite{zhang2021DPIR}.}
\label{fig:denoising_color_result}
\end{figure*}

\subsection{Results on Image Denoising}
We show grayscale and color image denoising results in Table~\ref{tab:denoising_grayscale_results} and Table~\ref{tab:denoising_color_results}, respectively. Compared methods include traditional models BM3D~\cite{dabov2007bm3d} and WNNM~\cite{gu2014weighted}, CNN-based models DnCNN~\cite{zhang2017DnCNN}, IRCNN~\cite{zhang2017IRCNN}, FFDNet~\cite{zhang2018ffdnet}, N3Net~\cite{plotz2018N3Net}, NLRN~\cite{liu2018NLRN}, FOCNet~\cite{jia2019focnet}, RNAN~\cite{zhang2019RNAN}, MWCNN~\cite{liu2018MWCNN} and DRUNet~\cite{zhang2021DPIR}. Following \cite{zhang2017DnCNN, zhang2021DPIR}, the compared noise levels include 15, 25 and 50. As one can see, our model achieves better performance than all compared methods. In particular, it surpasses the state-of-the-art model DRUNet by up to 0.3dB on the large Urban100 dataset that has 100 high-resolution testing images. 
It is worth pointing out that SwinIR only has 12.0M parameters, whereas DRUNet has 32.7M parameters. This indicates that the SwinIR architecture is highly efficient in learning feature representations for restoration. The visual comparison for grayscale and color image denoising of different methods are shown in Figs.~\ref{fig:denoising_gray_result} and~\ref{fig:denoising_color_result}. As we can see, our method can remove heavy noise corruption and preserve high-frequency image details, resulting in sharper edges and more natural textures. By contrast, other methods suffer from either over-smoothness or over-sharpness, and cannot recover rich textures.

\section{Conclusion}
In this paper, we propose a Swin Transformer-based image restoration model SwinIR. The model is composed of three parts: shallow feature extraction, deep feature extraction and HR reconstruction modules. In particular, we use a stack of residual Swin Transformer blocks (RSTB) for deep feature extraction, and each RSTB is composed of Swin Transformer layers, convolution layer and a residual connection. Extensive experiments show that SwinIR achieves state-of-the-art performance on three representative image restoration tasks and six different settings: classic image SR, lightweight image SR, real-world image SR, grayscale image denoising, color image denoising and JPEG compression artifact reduction, which demonstrates the effectiveness and generalizability of the proposed SwinIR. In the future, we will extend the model to other restoration tasks such as image deblurring and deraining.

\noindent\textbf{Acknowledgements}~~ This paper was partially supported by the ETH Zurich Fund (OK), a Huawei Technologies Oy (Finland) project, the China Scholarship Council and an Amazon AWS grant. Special thanks goes to Yijue Chen.

{\small
\bibliographystyle{ieee_fullname}
\bibliography{superresolution.bib}
}

\end{document}